\documentclass[a4paper,10pt]{article}
\usepackage{amssymb,amsmath,amsfonts}
\usepackage{fullpage}
\usepackage{rotating} 
\usepackage[latin1]{inputenc}
\begin{document}
\title{Improved estimators for a general class of beta regression models}
\date{}
\author{Alexandre B. Simas$^{a,}$\footnote{Corresponding author. E-mail: alesimas@impa.br},~ Wagner Barreto-Souza$^{b,}$\footnote{E-mail: wagnerbs85@hotmail.com}~ and Andréa V. Rocha$^{b,}$\footnote{E-mail: dearocha@yahoo.com}\\\\
\centerline{\small{
$^a$Associa\c{c}\~ao Instituto Nacional de Matem\'atica Pura e Aplicada, IMPA,}}\\
\centerline{\small{
Estrada D. Castorina, 110, Jd. Bot\^anico, 22460-320, Rio de Janeiro-RJ, Brasil}}\\
\centerline{\small{
$^b$Departamento de Estat\' \i stica, 
Universidade Federal de Pernambuco,
}}\\
\centerline{\small{
Cidade Universitária, 50740-540, Recife-PE, Brasil}}}
\sloppy
\maketitle
\begin{abstract}
In this paper we consider an extension of the beta regression model proposed by Ferrari and Cribari-Neto (2004). We
extend their model in two different ways, first, we let the regression structure be nonlinear, second, we allow a regression
structure for the precision parameter, moreover, this regression structure may also be nonlinear. Generally,
the beta regression is useful to situations where the response is restricted to the standard unit interval and the regression structure
involves regressors and unknown parameters. We derive general formulae for second-order biases of the maximum likelihood
estimators and use them to define bias-corrected estimators. Our formulae generalizes the results obtained by Ospina et al. (2006),
and are easily implemented by means of supplementary weighted linear regressions. We also compare these bias-corrected estimators
with three different estimators which are also bias-free to the second-order, one analytical and the other two based on bootstrap methods.
These estimators are compared by simulation. We present an empirical application.
\end{abstract}
\section{Introduction}

The beta distribution is a very flexible distribution and thus is commonly used to model data restricted to some
open interval on the line. The application turns to be more interesting when the interval which is being used is the
standard unit interval, $(0,1)$, since the data can be interpreted as rates or proportions. To work with this distribution
in a regression manner, several models were defined, see for instance, Ferrari and Cribari-Neto (2004), Kieschnick and McCullough
(2003), Paolino (2001), Vasconcellos and Cribari-Neto (2005), among others. But the one we will use here is an extension
of the one defined by Ferrari and Cribari-Neto (2004), mainly because this model is similar to the well known class of generalized linear
models (McCullagh and Nelder, 1989). Our objective in this paper is to reduce the bias of the maximum likelihood estimators (MLEs) for
this extended class of beta regression models.

We extend the model defined by Ferrari and Cribari-Neto (2004) in two ways, namely: (i) We allow the regression structure
to be nonlinear, such as the exponential family nonlinear models (Cordeiro and Paula, 1989) extend the generalized linear models. (ii)
We allow a regression structure on the precision parameter, this extension is similar to the way the generalized linear models
with dispersion covariates (e.g., Botter and Cordeiro, 1998) extend the generalized linear models. Further, we allow
the regression structure on the precision parameter to be nonlinear, an immediate consequence is that we are able to model heteroscedasticity in
a natural way, by means of this regression structure. Furthermore, we do not place any restrictions about the dispersion covariates, in the sense
that it can be, and usually is, a subset of the covariates of the mean. Moreover, if we see the section on numerical results
and the section on application to real data of Ospina et al. (2006), we are able to see that the precision parameter, usually has a large variance, thus is not sharply
estimated, we then expect the dispersion covariates to circumvent this problem.

The problem of modeling variances has been largely discussed in the statistical literature particularly in the econometric area (see, for instance, Park, 1966; Harvey, 1976).
Under normal errors, for instance, Cook and Weisberg (1983) and Atkinson (1985) present some graphical methods to detect heteroscedasticity,
Carrol and Ruppert (1988) develop diagnostic procedures using local influence methods for the variance parameter estimates in various nonlinear models
for the mean, whereas Verbyla (1993) compares full and residual maximum likelihood estimates based on case deletion and likelihood displacement. Moving
away from normal errors, Smyth (1989) describes a method which allows modeling the dispersion parameter in some generalized linear models whereas
Botter and Cordeiro (1997) present Bartlett correction expressions to improve likelihood ratio test in generalized linear models with dispersion
covariates. Barroso et al. (2002) present expressions to improve score tests in heteroscedastic Student-$t$ models and more recently Taylor and Verbyla (2004)
propose a joint modeling of location ans scale parameters in Student-$t$ models. Finally, Cysneiros et al. (2007) consider heteroscedastic linear models
with symmetric errors and discusses the diagnostic aspects of these models.

It is known that MLEs in nonlinear regression models are generally biased. These bias become
problematic when the study is being done in small samples. Bias does not pose a serious problem when the sample size $n$
is large, since its order is typically ${\cal O}(n^{-1})$, whereas the asymptotic standard error has order ${\cal O}(n^{-1/2})$.
Several authors have explored bias in regression models. Pike et al. (1979) investigated the magnitude of the bias in
unconditional estimates from logistic linear models, when the number of strata is large. Ratkwosky (1983) uses various
examples of normal nonlinear models to relate bias to the parameter effects curvature. Also for this class of models,
Cook et al. (1986) show that the bias may be due to the explanatory variables position in the sample space. Cordeiro
and McCullagh (1991) gave a general bias formulae in matrix notation for generalized linear models. 

Bias corrected versions of the MLEs of the parameters that index the beta distribution were obtained by Cordeiro et al. (1997)
and Cribari-Neto and Vasconcellos (2002). However, such results do not hold for models with regression structures.
Ospina et al. (2006) obtained expressions for the ${\cal O}(n^{-1})$ bias of the parameters of the beta regression models (Ferrari and Cribari-Neto, 2004).

The method used to obtain expressions of the ${\cal O}(n^{-1})$ bias of the parameters of this class of beta regression
models is the one given by Cox and Snell (1968). Further, an alternative approach was proposed by Firth (1993). He
suggested a \emph{preventive} method of bias reduction by modifying the score function prior to obtaining the parameter
estimates. It is also possible to perform bias adjustment using the estimated bias from a bootstrap resampling scheme,
which requires no explicit derivation of the bias function.

The chief goal of this paper is to obtain closed-form expressions for the second order biases of the MLEs
of the parameters, of the means of the responses, and of the precision parameters of the model.
The results are used to define bias corrected estimators to order ${\cal O}(n^{-1})$. We also consider bootstrap bias
adjustment and the bias adjustment given by Firth (1993).

The rest of this paper unfolds as follows. In Section 2, we introduce the general class of beta regression of interest
along with the score function and Fisher's information matrix. In Section 3, we derive a matrix expression for the second order biases of the MLEs of the parameters, and
consider analytical (corrective and preventive) and bootstrap bias correction schemes. We also show how the
biases of the MLEs of the parameters can be easily computed by means of auxiliary weighted linear regressions.
In Section 4, we obtain the second order biases of the MLEs of the means of the responses and precision parameters
of the model.  In Section 5, we consider some special cases in detail. In Section 6, we present simulation results that show that the proposed estimators have better performance
in small samples, in terms of bias, than the original MLEs. In Section 7,
we consider an empirical example. Finally, the paper is concluded in Section 8 with some final remarks.

\section{The model}

We say that a random variable $Y$ follows a beta distribution with parameters $p,q>0$, denoted by $B(p,q)$, if the
distribution of $Y$ admits the following density with respect to the Lebesgue measure:
\begin{equation}\label{dens1}
f(y;p,q) = \frac{\Gamma(p+q)}{\Gamma(p)\Gamma(q)} y^{p-1}(1-y)^{q-1},\quad y\in (0,1),
\end{equation}
where, $\Gamma(\cdot)$ is the gamma function. The mean and variance of $Y$ are, respectively
$$E(Y) = \frac{p}{p+q}\hbox{~~and~~} {\rm Var}(Y) = \frac{pq}{(p+q)^2(p+q+1)}.$$

Ferrari and Cribari-Neto (2004) defined a regression structure for beta distributed responses that differs from (\ref{dens1}).
Let $\mu = p/(p+q)$ and $\phi = p+q$, i.e., $p=\mu\phi$ and $q=(1-\mu)\phi$. Under this new parameterization, if
$Y\sim B(p,q)$, then $E(Y) = \mu$ and Var$(Y) = V(\mu)/(1+\phi)$, where $V(\mu) = \mu(1-\mu)$ denotes a ``variance function'',
under this parameterization, we will use the notation $Y\sim {\cal B}(\mu,\phi)$. We also note that this parameterization
was already known in the statistical literature (see, for instance, J\o rgensen, 1997, p. 33).
Further, $\phi$ plays the role of a precision parameter, in the sense that, for fixed $\mu$, the larger the $\phi$,
the smaller the variance of the response. Using this new parameterization, the beta density in (\ref{dens1}) can be
written as
\begin{equation}\label{densmuphi}
f(y;\mu,\phi) = \frac{\Gamma(\phi)}{\Gamma(\mu\phi)\Gamma((1-\mu)\phi)} y^{\mu\phi -1} (1-y)^{(1-\mu)\phi-1},\quad y\in(0,1),
\end{equation}
and the log-density is thus
$$\log f(y;\mu,\phi) = \log \Gamma(\phi) - \log \Gamma(\mu\phi) - \log \Gamma((1-\mu)\phi) + (\mu\phi)\log y + \{(1-\mu)\phi -1\}\log(1-y),$$
with, $0<\mu<1$ and $\phi>0$, since $p,q>0$.

Let $y = (y_1,\ldots,y_n)^T$ be a random sample, where $y_i \sim {\cal B}(\mu_i,\phi_i)$, $i=1,\ldots,n$. Suppose
the mean and the precision parameter of $y_i$ satisfies the following functional relations:
\begin{equation}\label{regr}
g_1(\mu_i) = \eta_{1i} = f_1 (x_i^T;\beta)\hbox{~~and~~}g_2(\phi_i) = \eta_{2i} = f_2(z_i^T;\theta),
\end{equation}
where $\beta = (\beta_1,\ldots,\beta_k)^T$ and $\theta = (\theta_1,\ldots,\theta_h)^T$ are vectors of unknown regression
parameters which are assumed to be functionally independent, $\beta\in\mathbb{R}^k$ and $\theta\in\mathbb{R}^h$, $k+h<n$, $\eta_{1i}$ and $\eta_{2i}$ are predictors,
and $x_{i1},\ldots,x_{iq_1},z_{i1},\ldots,z_{iq_2}$ are observations on $q_1$ and $q_2$ \emph{known} covariates, which need not to be exclusive.
We assume that the derivative matrices $\tilde{X} = \partial\eta_1/\partial\beta$ and $\tilde{Z} = \partial\eta_2/\partial\theta$
have rank $k$ and $h$, respectively. Moreover, we assume that the link functions $g_1:(0,1)\to\mathbb{R}$ and
$g_2:(0,\infty)\to\mathbb{R}$ are
strictly monotonic and twice differentiable. A number of different link functions can be used, such as the logit
specification $g_1 (\mu) = \log\{\mu/(1-\mu)\}$, the probit function $g_1(\mu) = \Phi^{-1}(\mu)$, where
$\Phi(\cdot)$ denotes the standard normal distribution function, the complementary log-log function $g_1(\mu) = \log\{-\log(1-\mu)\}$,
among others, and for $g_2$, $g_2(\phi) = \log \phi$, the logarithmic function, $g_2(\phi) = \sqrt{\phi}$, the square root function,
$g_2(\phi) = \phi$ (with special attention on the positivity of the estimates), among others. A rich discussion of link functions can be found in McCullagh and Nelder (1989); see also
Atkinson (1985, Chapter 7).

The log-likelihood function for this class of beta regression models has the form
\begin{equation}\label{loglik}
\ell(\beta,\theta) = \sum_{i=1}^n \ell_i(\mu_i,\phi_i),
\end{equation}
where
\begin{eqnarray*}
\ell_i(\mu_i,\phi_i) &=&\log\Gamma(\phi_i) - \log\Gamma((1-\mu_i)\phi_i)+(\mu_i\phi_i-1)\log y_i\\
&&+\{(1-\mu_i)\phi_i-1\}\log(1-y_i);
\end{eqnarray*}
$\mu_i = g_1^{-1}(\eta_{1i}), \phi_i = g_2^{-1}(\eta_{2i})$, as defined in (\ref{regr}), are functions of $\beta$
and $\theta$, respectively. It is possible to show that this beta regression model is regular, in the sense
that all the regularity conditions described in Cox and Hinkley (1974, p. 107) hold. It is also possible to show
that the MLEs are unique.

The components of the score vector, obtained by differentiation of the log-likelihood function with respect to the parameters,
are given, for $r=1,\ldots,k$, as
$$U_r(\beta,\theta) = \frac{\partial \ell(\beta,\theta)}{\partial\beta_r} = \sum_{i=1}^n \phi_i(y_i^\ast - \mu_i^\ast)\frac{d\mu_i}{d\eta_{1i}}\frac{\partial\eta_{1i}}{\partial\beta_k},$$
where $d\mu_i/d\eta_{1i} = 1/g_1'(\mu_i)$, $y_i^\ast = \log(y_i/(1-y_i))$, $\mu_i^\ast = \psi(\mu_i\phi_i)-\psi((1-\mu_i)\phi_i)$, and $\psi(\cdot)$ is the digamma\footnote{We denote generally the polygamma function by
$\psi^{(m)}(\cdot)$, $m=0,1,\ldots,$ where $\psi^{(m)}(x) = \left( d^{m+1}/dx^{m+1}\right)\log\Gamma(x),x>0.$} function, together with
$$U_R(\beta,\theta) = \frac{\partial\ell(\beta,\theta)}{\partial\theta_R}= \sum_{i=1}^n \left\{\mu_i(y_i^\ast - \mu_i^\ast) +\psi(\phi_i) - \psi((1-\mu_i)\phi_i) + \log(1-y_i)\right\}\frac{d\phi_i}{d\eta_{2i}}\frac{\partial\eta_{2i}}{\partial\theta_R},$$
where $d\phi_i/d\eta_{2i} = 1/g_2'(\phi_i)$, and $R=1,\ldots,h$. Further, the regularity conditions implies that
$$E\left(\log\frac{y_i}{1-y_i}\right) = \psi(\mu_i\phi_i) - \psi((1-\mu_i)\phi_i),$$
and
$$E\{\log(1-y_i)\} = \psi((1-\mu_i)\phi_i) - \psi(\phi_i).$$

Consider the complete parameter vector $\zeta = (\beta^T,\theta^T)^T$. Define the vectors $y^\ast = (y_1^\ast,\ldots,y_n^\ast)^T$,
$\mu^\ast = (\mu_1^\ast,\ldots,\mu_n^\ast)^T$, $v = (v_1,\ldots,v_n)^T$, the matrix \label{formulat1t2} $T_1 = {\rm diag}(d\mu_i/d\eta_{1i}), T_2 = {\rm diag}(d\phi_i/d\eta_{2i})$, $\Phi = {\rm diag}(\phi_i)$,
with diag$(\mu_i)$ denoting the $n\times n$ diagonal matrix with typical element $\mu_i,i=1,\ldots,n$, and where
$v_i = \mu_i(y_i^\ast - \mu_i^\ast) +\psi(\phi_i) - \psi((1-\mu_i)\phi_i) + \log(1-y_i)$. Therefore, we can write the
$(k+h)\times 1$ dimensional score vector $U(\zeta)$ in the form $(U_\beta(\beta,\theta)^T,U_\theta(\beta,\theta)^T)^T$,
with
\begin{equation}\label{scorevec}
\begin{array}{ccl}
U_\beta(\beta,\theta) &=& \tilde{X}^T \Phi T_1 (y^\ast-\mu^\ast),\\
U_\theta(\beta,\theta) &=& \tilde{Z}^T T_2 v.
\end{array}
\end{equation}
The MLEs of $\beta$ and $\theta$ are obtained as the solution of the nonlinear system
$U(\zeta) = 0$. In practice, the MLEs can be obtained through a numerical maximization of
the log-likelihood function using a nonlinear optimization algorithm, e.g., BFGS. For details, see Press et al. (1992).

Define $P$ as the $2n\times (k+h)$ dimensional matrix
\begin{equation}\label{matrizp}
P = \left(\begin{array}{cc}
\tilde{X} & 0\\
0 & \tilde{Z}
\end{array}\right).
\end{equation}
Also, let $W$ be the $2n\times 2n$ matrix
\begin{equation}\label{matrizw}
W = \left(\begin{array}{cc}
W_{\beta\beta} & W_{\beta\theta}\\
W_{\beta\theta} & W_{\theta\theta}
\end{array}\right),
\end{equation}
with
\begin{eqnarray*}
W_{\beta\beta} &=& {\rm diag}\left(\phi_i^2 a_i \left(\frac{d\mu_i}{d\eta_{1i}}\right)^2\right),\\
W_{\beta\theta} &=& {\rm diag} \left(\phi_i\{\mu_i a_i - \psi'((1-\mu_i)\phi_i)\}\left(\frac{d\mu_i}{d\eta_{1i}}\right)\left(\frac{d\phi_i}{d\eta_{2i}}\right)\right),\\
W_{\theta\theta} &=& {\rm diag} \left(b_i \left(\frac{d\phi_i}{d\eta_{2i}}\right)^2\right).
\end{eqnarray*}
Here, $a_i = \psi'((1-\mu_i)\phi_i) + \psi'(\mu_i\phi_i)$ and $b_i = \psi'((1-\mu_i)\phi_i)(1-\mu_i)^2 + \psi'(\mu_i\phi_i)\mu_i^2 - \psi'(\phi_i)$.
Now, using (\ref{matrizp}) and (\ref{matrizw}), it is possible to obtain Fisher's information matrix for the parameter
vector $\zeta = (\beta^T,\theta^T)^T$ as
$$K(\zeta) = P^T W P.$$

Note that $W_{\beta\theta} \neq 0$, thus indicating that the parameters $\beta$ and $\theta$ are not orthogonal, in constrast
to the class of generalized linear models (McCullagh and Nelder, 1989), where such orthogonality holds. Nevertheless, the MLEs $\hat{\zeta}$ and
$K(\hat{\zeta})$ are consistent estimators of $\zeta$ and $K(\zeta)$, respectively, where $K(\hat{\zeta})$ is the Fisher's information matrix
evaluated at $\hat{\zeta}$. Assuming that $J(\zeta) = \lim_{n\to\infty} K(\zeta)/n$ exists and is nonsingular, we have that $\sqrt{n}\left(\hat{\zeta}-\zeta\right)\stackrel{d}{\rightarrow} N_{k+h}(0,J(\zeta)^{-1}),$
where, $\stackrel{d}{\rightarrow}$ denotes convergence in distribution. Hence if $\zeta_r$ denotes the $r$th component of $\zeta$, it follows that
$$\left(\hat{\zeta}-\zeta\right) \{K(\hat{\zeta})^{rr}\}^{-1/2} \stackrel{d}{\rightarrow} N(0,1),$$
where $K(\hat{\zeta})^{rr}$ is the $r$th diagonal element of $K(\hat{\zeta})^{-1}$. Let $K(\hat{\zeta})^{RR}$ be the $(k+R)$th diagonal element of $K(\hat{\zeta})^{-1}$.
Then, if $0<\alpha<1/2$, and $q_\gamma$ represents the $\gamma$ quantile of the $N(0,1)$ distribution, we have, for $r=1,\ldots,k$, $\hat{\beta}_r \pm q_{1-\alpha/2}\left( K(\hat{\zeta})^{rr}\right)^{1/2}$
and $\hat{\theta}_R \pm q_{1-\alpha/2}\left( K(\hat{\zeta})^{RR}\right)^{1/2}$ as the limits of asymptotic confidence intervals for $\beta_r$ and $\theta_R$,
respectively, both with asymptotic coverage of $100(1-\alpha)\%$. The asymptotic variances of $\hat{\beta}_r$ and $\hat{\theta}_R$ are estimated by
$K(\hat{\zeta})^{rr}$ and $K(\hat{\zeta})^{RR}$, respectively.

\section{Bias correction of the MLEs of $\beta$ and $\theta$}\label{biasbetatheta}

We begin by obtaining an expression for the second order biases of the MLEs of $\beta$ and $\theta$ in this general class of beta regression models
using Cox and Snell's (1968) general formula. With this expression we will be able to obtain bias corrected estimates of the unknown parameters.

We now introduce the following total \label{cumulants} log-likelihood derivatives in which we reserve lower-case subscripts $r,s,t,u,\ldots$ to denote components
of the $\beta$ vector and upper-case subscripts $R,S,T,U,\ldots$ for components of the $\theta$ vector: $U_r = \partial \ell/\partial \beta_r$,
$U_{rS} = \partial^2\ell/\partial\beta_r\theta_S$, $U_{rsT} = \partial^3\ell/\partial\beta_r\partial\beta_s\partial\theta_T$, and so on. The standard
notation will be adopted for the moments of the log-likelihood derivatives (Lawley, 1956): $\kappa_{rs} = E(U_{rs})$, $\kappa_{r,s} = E(U_rU_s)$,
$\kappa_{rs,T} = E(U_{rs}U_T)$, etc., where all $\kappa$'s to a total over sample and are, in general, of order ${\cal O}(n)$. We define the derivatives
of the moments by $\kappa_{rs}^{(t)} = \partial \kappa_{rs}/\partial\beta_t$, $\kappa_{rs}^{(T)} = \partial\kappa_{rs}/\partial\theta_T$, etc. Not
all the $\kappa$'s are functionally independent. For example, $\kappa_{rs,t} = \kappa_{rst} - \kappa_{rs}^{(t)}$ gives the covariance between the first
derivative of $\ell(\beta,\theta)$ with respect to $\beta_t$ and the mixed second derivative with respect to $\beta_r,\beta_s$. Further,
let $\kappa^{r,s} = - \kappa^{rs}$, $\kappa^{R,s} = -\kappa^{Rs}$, $\kappa^{r,S} = -\kappa^{rS}$ and $\kappa^{R,S} = -\kappa^{RS}$ be typical elements
of $K(\zeta)^{-1}$, the inverse of the Fisher's information matrix, which are ${\cal O}(n^{-1})$.

Let $B(\hat{\zeta}_a)$ be the ${\cal O}(n^{-1})$ bias of the MLE for the $a$th component of the parameter vector $\hat{\zeta} = (\hat{\zeta}_1,\ldots,\hat{\zeta}_k,\hat{\zeta}_{k+1},\ldots,\hat{\zeta}_{k+h}) = (\hat{\beta}^T,\hat{\theta}^T)^T$.
From the general expression for the multiparameter ${\cal O}(n^{-1})$ biases of the MLEs given by Cox and Snell (1968), we can write
\begin{eqnarray}
B(\hat{\zeta}_a) &=& \sum_{r,s,u} \kappa^{ar} \kappa^{su} \left\{\kappa_{rs}^{(u)}-\frac{1}{2}\kappa_{rsu}\right\} + \sum_{R,s,u} \kappa^{aR} \kappa^{su} \left\{\kappa_{Rs}^{(u)}-\frac{1}{2}\kappa_{Rsu}\right\}\nonumber\\
&+& \sum_{r,S,u} \kappa^{ar} \kappa^{Su} \left\{\kappa_{rS}^{(u)}-\frac{1}{2}\kappa_{rSu}\right\} + \sum_{r,s,U} \kappa^{ar} \kappa^{sU} \left\{\kappa_{rs}^{(U)}-\frac{1}{2}\kappa_{rsU}\right\}\nonumber\\
&+& \sum_{R,S,u} \kappa^{aR} \kappa^{Su} \left\{\kappa_{RS}^{(u)}-\frac{1}{2}\kappa_{RSu}\right\} + \sum_{R,s,U} \kappa^{aR} \kappa^{sU} \left\{\kappa_{Rs}^{(U)}-\frac{1}{2}\kappa_{RsU}\right\}\nonumber\\
&+& \sum_{r,S,U} \kappa^{ar} \kappa^{SU} \left\{\kappa_{rS}^{(U)}-\frac{1}{2}\kappa_{rSU}\right\} + \sum_{R,S,U} \kappa^{aR} \kappa^{SU} \left\{\kappa_{RS}^{(U)}-\frac{1}{2}\kappa_{RSU}\right\}\label{coxsnell}
\end{eqnarray}
From (\ref{matrizw}) we note that the entries of the matrix $W_{\beta\theta}$ are not all zero, which makes the derivation cumbersome, since all
terms in (\ref{coxsnell}) must be considered. These terms together with the cumulants needed to obtain them are given in the Appendix. After some
tedious algebra, we arrive at the following expression, in matrix form, for the second order bias of $\hat{\beta}$:
\begin{eqnarray*}
B(\hat{\beta}) &=& K^{\beta\beta}\tilde{X}^T\left[M_1 P_{\beta\beta} + (M_2+M_3)P_{\beta\theta}+M_5 P_{\theta\theta}\right] + K^{\beta\beta}\tilde{X}^T\left[N_2 G - N_1 F\right]\\
&+& K^{\beta\theta} \tilde{Z}^T\left[M_2  P_{\beta\beta} + (M_4+M_5)P_{\beta\theta}+M_6 P_{\theta\theta}\right]+ K^{\beta\theta} \tilde{Z}^T\left[N_2 F-N_3 G \right],
\end{eqnarray*}
where $M_1$ to $M_6$ and $N_1$ to $N_3$ are given in (\ref{matrizm1})-(\ref{matrizn3}) in the Appendix, $K^{\beta\beta}$, $K^{\beta\theta}$ and $K^{\theta\theta}$ are the matrices
formed by the $(r,s)$th, $(r,S)$th and $(R,S)$th elements of the inverse of the Fisher's information matrix with $r,s=1,\ldots,k$, $R,S=k+1,\ldots,k+h$,
respectively, $P_{\beta\beta}$, $P_{\beta\theta}$ and $P_{\theta\theta}$ are the $n\times 1$ dimensional vectors containing the diagonal elements
of $\tilde{X}K^{\beta\beta}\tilde{X}^T$,\label{formulas} $\tilde{X}K^{\beta\theta}\tilde{Z}^T$ and $\tilde{Z}K^{\theta\theta}\tilde{Z}^T$, respectively,
$F={\rm diag}(F_1,\ldots,F_n)\mathbf{1}$, $F_i = {\rm tr}(\tilde{X}_i K^{\beta\beta})$, $\tilde{X}_i$ is a $k\times k$ matrix with elements $\partial^2\eta_{1i}/\partial\beta_r\beta_s$,
$\mathbf{1}$ is an $n\times 1$ vector of ones, $G = {\rm diag}(G_1,\ldots,G_n)\mathbf{1}$, $G_i = {\rm tr}(\tilde{Z}_i K^{\theta\theta})$, $\tilde{Z}_i$ is a $h\times h$ matrix with elements $\partial^2\eta_{2i}/\partial\theta_R\partial\theta_S$.

We define the $2n\times 1$-vectors $\omega_1$ and $\omega_2$ as
$$\omega_1 = \left(\begin{array}{c}
M_1 P_{\beta\beta} + (M_2+M_3)P_{\beta\theta}+M_5 P_{\theta\theta}\\
M_2  P_{\beta\beta} + (M_4+M_5)P_{\beta\theta}+M_6 P_{\theta\theta}
\end{array}\right)$$
and
$$\omega_2 = \left(\begin{array}{c}
N_2 G - N_1 F\\
 N_2 F -N_3 G
\end{array}\right).$$
We also consider the $k\times (k+h)$ upper block of the matrix $K(\zeta)^{-1}$ given by
$$K^{\beta\ast} = \left(K^{\beta\beta} K^{\beta\theta}\right).$$
The ${\cal O}(n^{-1})$ bias of $\hat{\beta}$ can now be written as
\begin{equation}\label{biasbeta}
B(\hat{\beta}) = K^{\beta\ast} P^T (\omega_1 + \omega_2).
\end{equation}
We now turn to the ${\cal O}(n^{-1})$ bias of $\hat{\theta}$. Analogously, we have the following expression, in matrix form, for the second order bias of $\hat{\theta}$:
\begin{eqnarray*}
B(\hat{\theta}) &=& K^{\beta\theta}\tilde{X}^T\left[M_1 P_{\beta\beta} + (M_2+M_3)P_{\beta\theta}+M_5 P_{\theta\theta}\right] + K^{\beta\theta}\tilde{X}^T\left[N_2 G - N_1 F\right]\\
&+& K^{\theta\theta} \tilde{Z}^T\left[M_2  P_{\beta\beta} + (M_4+M_5)P_{\beta\theta}+M_6 P_{\theta\theta}\right]+ K^{\theta\theta} \tilde{Z}^T\left[N_2 F -N_3 G \right],
\end{eqnarray*}
Then, considering the $h\times (k+h)$ lower block of the matrix $K(\zeta)^{-1}$, given by
$$K^{\theta\ast} = \left(K^{\beta\theta}K^{\theta\theta}\right),$$
we can write the ${\cal O}(n^{-1})$ bias of $\hat{\theta}$ as
\begin{equation}\label{biastheta}
B(\hat{\theta}) = K^{\theta\ast} P^T (\omega_1 + \omega_2).
\end{equation}
Thus, combining (\ref{biasbeta}) and (\ref{biastheta}), we are able to find that the ${\cal O}(n^{-1})$ bias of the MLE of the joint vector $\zeta = (\beta^T,\theta^T)^T$ can be written, in matrix form, as
$$B(\hat{\zeta}) = K(\zeta)^{-1} P^T(\omega_1 + \omega_2) = (P^T W P)^{-1} P^T(\omega_1 + \omega_2).$$
Now, let $\xi_1 = W^{-1}\omega_1$ and $\xi_2 = W^{-1}\omega_2$, then, the previous expression becomes
\begin{equation}\label{biaszeta}
B(\hat{\zeta}) = \left(P^T W P\right)^{-1}P^T W(\xi_1 + \xi_2).
\end{equation}
Thus, the ${\cal O}(n^{-1})$ bias of $\hat{\zeta}$ (\ref{biaszeta}) is easily obtained as the vector of regression coefficients in the formal weighted
linear regression of $\xi = \xi_1 + \xi_2$ on the columns of $P$ with $W$ as weight matrix.

The ${\cal O}(n^{-1})$ bias (\ref{biaszeta}) is expressed as the sum of two quantities: (i) $B_1 = \left(P^T W P\right)^{-1}P^T W\xi_1$, the bias for a linear
beta regression model with dispersion covariates with model matrices $\tilde{X}$ and $\tilde{Z}$, and thus generalizes the expression obtained by Ospina et al. (2006),
and (ii) an additional quantity $B_2 = \left(P^T W P\right)^{-1}P^T W\xi_2$ due to the nonlinearity of the functions $f_1(x_i;\beta)$ and $f_2(z_i;\theta)$, and which vanishes
if both $f_1$ and $f_2$ are linear in $\beta$ and $\theta$, respectively.

Now we can define our first bias-corrected estimator $\tilde{\zeta}$ as
$$\tilde{\zeta} = \hat{\zeta} - \hat{B}(\hat{\zeta}),$$
where $\hat{B}(\hat{\zeta})$ denotes the MLE of $B(\hat{\zeta})$, that is, the unknown parameters are replaced by their MLEs. Since the bias $B(\hat{\zeta})$
is of order ${\cal O}(n^{-1})$, it is not difficult to show that the asymptotic normality $\sqrt{n}\left(\tilde{\zeta}-\zeta\right)\stackrel{d}{\rightarrow}N_{k+h}(0,J^{-1}(\zeta))$
still holds, where, as before, we assume that $J(\zeta) = \lim_{n\to\infty} K(\zeta)/n$ exists and is nonsingular. From the asymptotic normality
of $\tilde{\zeta}$, we have that $\tilde{\zeta}_a \pm q_{1-\alpha/2}\left\{K(\tilde{\zeta})^{aa}\right\}^{1/2}$, for $a=1,\ldots,k,k+1,\ldots,k+h$.
The asymptotic variance of $\tilde{\zeta}_a$ is estimated by $K(\tilde{\zeta})^{aa}$, where $K(\tilde{\zeta})^{aa}$ is the $a$th diagonal element of
the inverse of the Fisher's information matrix evaluated at $\tilde{\zeta}$.

We now turn to the bias-correction approach developed by Firth (1993) called the ``preventive'' method. This method also remove the ${\cal O}(n^{-1})$ bias
and consists of modifying the original score function:
$$U^\ast(\zeta) = U(\zeta) + K(\zeta)B(\zeta),$$
where $K(\zeta)$ is the information matrix and $B(\zeta)$ is the ${\cal O}(n^{-1})$ bias. The solution $\check{\zeta}$ of $U^\ast = 0$
is a bias-corrected estimator,
to order ${\cal O}(n^{-1})$, for $\zeta$. For our general class of beta regression models, the substitution of $B(\zeta)$ by (\ref{biaszeta}) yields
the following form for the modified score function:
\begin{equation}\label{modscore}
U^\ast(\zeta) = U(\zeta) + P^T(\omega_1 + \omega_2).
\end{equation}

The estimator $\check{\zeta}$, obtained as the root of the modified score function in (\ref{modscore}), is consistent and asymptotically normal: $\sqrt{n}\left(\check{\zeta}-\zeta\right)\stackrel{d}{\rightarrow}N_{k+h}(0,J^{-1}(\zeta))$,
with $J(\zeta)$ as given before. We also have that $\check{\zeta}_a \pm q_{1-\alpha/2}\left\{K(\check{\zeta})^{aa}\right\}^{1/2}$, for $a=1,\ldots,k,k+1,\ldots,k+h$.
The asymptotic variance of $\check{\zeta}_a$ is estimated by $K(\check{\zeta})^{aa}$, where $K(\check{\zeta})^{aa}$ is the $a$th diagonal element of
the inverse of the Fisher's information matrix evaluated at $\check{\zeta}$.

A third, and the last approach we consider here, to bias-correcting MLEs of the regression parameters is based upon the numerical estimation of the bias
through the bootstrap resampling scheme introduced by Efron (1979).  Let
${\bf y} = (y_1,\ldots, y_n)^{\top}$ be a random sample of size $n$, where each element is a random draw from the random
variable $Y$ which has the distribution function $F=F(\zeta)$. Here, $\zeta$ is the parameter that indexes the distribution, and
is viewed as a functional of $F$, i.e., $\zeta = t(F)$. Finally, let $\hat{\zeta}$ be an estimator of $\zeta$ based on ${\bf y}$; we write $\hat{\zeta} = s({\bf y})$.

The application of the bootstrap method consists in obtaining, from the original sample ${\bf y}$, a large number
of pseudo-samples ${\bf y}^{*} = (y_{1}^{*},\ldots, y_{n}^{*})^{\top}$, and then extracting information from these samples to improve
inference. Bootstrap methods can be classified into two classes, depending on how the sampling is performed: parametric and nonparametric.
In the parametric version, the bootstrap samples are obtained from $F(\hat{\zeta})$, which we shall denote as $F_{\hat{\zeta}}$, whereas in the nonparametric version they are
obtained from the empirical distribution function $\hat{F}$, through sampling with replacement. Note that the nonparametric bootstrap does not entail parametric assumptions.

Let $B_{F}(\hat{\zeta},\zeta)$ be the bias of the estimator $\hat{\zeta} = s({\bf y})$, that 
is, 
\[B_{F}(\hat{\zeta}, \zeta) = {\rm E}_{F}[\hat{\zeta} - \zeta] = {\rm E}_{F}[s({\bf y})] - t(F),\]
where the subscript $F$ indicates that expectation is taken with respect to $F$. The bootstrap estimators of
the bias in the parametric and nonparametric versions are obtained by replacing the true distribution $F$, which generated the original
sample,
with $F_{\hat{\zeta}}$ and $\hat{F}$, respectively, in the above expression.
Therefore, the parametric and nonparametric estimates of the bias are given, respectively, by
\[B_{F_{\hat{\zeta}}}(\hat{\zeta}, \zeta) = {\rm E}_{F_{\hat{\zeta}}}[s({\bf y})] - t(F_{\hat{\zeta}})
  \quad{\rm and}\quad B_{\hat{F}}(\hat{\zeta}, \zeta) = {\rm E}_{\hat{F}}[s({\bf y})] - t(\hat{F}).\]

\noindent If $B$ bootstrap samples $({\bf y}^{*1}, {\bf y}^{*2}, \ldots, {\bf y}^{*B})$ are generated independently from the original
sample ${\bf y}$, and the respective boostrap replications $(\hat{\zeta}^{*1}, \hat{\zeta}^{*2}, \ldots, \hat{\zeta}^{*B})$
are calculated,
where $\hat{\zeta}^{*b} = s({\bf y}^{*b})$, $b = 1, 2, \ldots, B$, then it is possible to approximate the bootstrap expectations 
${\rm E}_{F_{\hat{\zeta}}}[s({\bf y})]$ and ${\rm E}_{\hat{F}}[s({\bf y})]$
by the mean $\hat{\zeta}^{*(\cdot)} = \frac{1}{B}\sum_{b=1}^{B}\hat{\zeta}^{*b}$.
Therefore, the bootstrap bias estimates based on $B$ replications of $\hat{\zeta}$ are
\begin{equation}\label{biasboot}
\hat{B}_{F_{\hat{\zeta}}}(\hat{\zeta}, \zeta) = \hat{\zeta}^{*(\cdot)} - s({\bf y})\quad{\rm and}\quad
\hat{B}_{\hat{F}}(\hat{\zeta}, \zeta) = \hat{\zeta}^{*(\cdot)} - s({\bf y}),
\end{equation}

\noindent for the parametric and nonparametric versions, respectively.

By using the two bootstrap bias estimates presented above, we arrive at the following two bias-corrected, to order ${\cal O}(n^{-1})$, estimators:
\begin{eqnarray*}
\overline{\zeta}_{1} &=& s({\bf y}) - \hat{B}_{F_{\hat{\zeta}}}(\hat{\zeta}, \zeta) = 2\hat{\zeta} - \hat{\zeta}^{*(\cdot)},\\
\overline{\zeta}_{2} &=& s({\bf y}) - \hat{B}_{\hat{F}}(\hat{\zeta}, \zeta) = 2\hat{\zeta} - \hat{\zeta}^{*(\cdot)}.
\end{eqnarray*}
The corrected estimates $\overline{\zeta}_{1}$ and $\overline{\zeta}_{2}$ were called 
constant-bias-correcting (CBC) estimates by MacKinnon and Smith (1998).

Since we are dealing with regression models and not with a random sample we need some minor modifications to the algorithm given above.

For the nonparametric case, assume we want to fit a regression model with response variable $y$ and predictors $x_1,\ldots,x_{q_1},z_1,\ldots,z_{q_2}$.
We have a sample of $n$ observations $p_i^T = (y_i,x_{i1},\ldots,x_{iq_1},z_{i1},\ldots,z_{iq_2})$, $i=1,\ldots,n$. Thus we use the nonparametric bootstrap
method described above to obtain $B$ bootstrap samples of the $p_i^T$, fit the model and save the coefficients from each bootstrap sample. We
can then obtain bias corrected estimates for the regression coefficients using the methods described above. This is the so-called Random-$x$ resampling.

For the parametric case, assume we have the same model as for the nonparametric case, we thus obtain the estimates $\hat{\mu}_i$ and $\hat{\phi}_i$
(such as in our case where the distribution is indexed by $\mu$ and $\phi$) and using the
parametric method described above, we obtain $B$ bootstrap samples for $\hat{y}_i$ from the distribution $F(\hat{\mu}_i,\hat{\phi}_i)$, $i=1,\ldots,n$. We would then
regress each set of bootstrapped values $y_b^{\ast}$ on the covariates $x_1,\ldots,x_{q_1},z_1,\ldots,z_{q_2}$ to obtain bootstrap replications of the regression
coefficients. We can, again, obtain bias corrected estimates for the regression coefficients using the methods described above. This method is called
Fixed-$x$ resampling.

\section{Bias correction of the MLEs of $\mu$ and $\phi$}\label{biasmuphi}

In this Section we obtain the results that are the most valuable to the practioners, namely, the ${\cal O}(n^{-1})$ bias of $\mu$ and of $\phi$,
since, for practioners, the interest in a data analysis relies on sharp estimates of the responses and of the precision parameters. The fact that
these results must be computed apart comes from the fact that if $\ddot{\beta}$ and $\ddot{\theta}$ are bias-free estimators, to order ${\cal O}(n^{-1})$, it is not
true, in general, that $\ddot{\mu}_i = g_1^{-1}( f_1(x_i;\ddot{\beta}))$ and $\ddot{\phi}_i = g_2^{-1}(f_2(z_i;\ddot{\theta}))$ will also be bias-free to
order ${\cal O}(n^{-1})$. Nevertheless, for practioners, it is even more important to correct the means of the responses and the precision parameters than
correcting the regression parameters. Moreover, the ${\cal O}(n^{-1})$ bias of $\hat{\mu}$ for the linear beta regression model was not presented in Ospina et al. (2006).

We shall first obtain the ${\cal O}(n^{-1})$ bias of the MLEs of $\eta_1$ and $\eta_2$. Using (\ref{regr}) we find, by Taylor expansion, that to order ${\cal O}(n^{-1})$:
$$f_1(x_i^T; \hat{\beta}) - f_1(x_i^T; \beta) = \nabla_\beta(\eta_{1i})^T(\hat{\beta}-\beta) + \frac{1}{2}(\hat{\beta}-\beta)^T \tilde{X}_i (\hat{\beta}-\beta),$$
and
$$f_2(z_i^T; \hat{\theta}) - f_2(z_i^T; \theta) = \nabla_\theta(\eta_{2i})^T (\hat{\theta}-\theta) + \frac{1}{2}(\hat{\theta}-\theta)^T\tilde{Z}_i (\hat{\theta}-\theta),$$
where $\nabla_\beta(\eta_{1i})$ is a $k\times 1$ vector with the derivatives $\partial\eta_{1i}/\partial\beta_r$, $\nabla_\theta(\eta_{2i})$ is a
$h\times 1$ vector with the derivatives $\partial\eta_{2i}/\partial\theta_R$, and all the other quantities were previously defined in page \pageref{formulas}.

Thus, taking expectations on both sides of the above expression yields to this order
$$B(\hat{\eta_1}) = \tilde{X}B(\hat{\beta}) + \frac{1}{2}F,$$
and
$$B(\hat{\eta_2}) = \tilde{Z}B(\hat{\theta}) + \frac{1}{2}G,$$
where, $F$ and $G$ were defined in page \pageref{formulas}, and since $K^{\beta\beta}$ and $K^{\theta\theta}$ are the asymptotic covariance matrices of
$\hat{\beta}$ and $\hat{\theta}$, respectively.

From similar calculations we obtain to order ${\cal O}(n^{-1})$
$$B(\hat{\mu}_i) = B(\hat{\eta}_{1i})\frac{d\mu_i}{d\eta_{1i}} + \frac{1}{2}{\rm Var}(\hat{\eta_{1i}})\frac{d^2\mu_i}{d\eta_{1i}^2}$$
and
$$B(\hat{\phi}_i) = B(\hat{\eta}_{2i})\frac{d\phi_i}{d\eta_{2i}} + \frac{1}{2}{\rm Var}(\hat{\eta_{2i}})\frac{d^2\mu_i}{d\eta_{2i}^2}.$$

Let $T_1$ and $T_2$ be as in page \pageref{formulat1t2}, further, let $S_1 = {\rm diag}(d^2\mu_i/d\eta_{1i}^2)$ and $S_2 = {\rm diag}(d^2\phi_i/d\eta_{2i}^2)$.
Then, we can write the above expressions in matrix notation as
\begin{equation}\label{biasmu}
B(\hat{\mu}) = \frac{1}{2}T_1(2\tilde{X}B(\hat{\beta})+F) + \frac{1}{2} S_1 P_{\beta\beta}
\end{equation}
and
\begin{equation}\label{biasphi}
B(\hat{\phi}) = \frac{1}{2}T_2(2\tilde{Z}B(\hat{\theta}) + G) + \frac{1}{2} S_2 P_{\theta\theta},
\end{equation}
where $P_{\beta\beta}$ and $P_{\theta\theta}$ were defined in page \pageref{formulas} and since the asymptotic covariance matrices of $\hat{\eta_1}$ and
$\hat{\eta_2}$ are $\tilde{X}K^{\beta\beta}\tilde{X}^T$ and $\tilde{Z}K^{\theta\theta}\tilde{Z}^T$, respectively.

If we combine (\ref{biasmu}) and (\ref{biasphi}) with (\ref{biasbeta}) and (\ref{biastheta}), we will have the following explicit expressions for the
${\cal O}(n^{-1})$ biases of $\hat{\mu}$ and $\hat{\phi}$, respectively:
$$B_1(\hat{\mu}) = \frac{1}{2}T_1(2\tilde{X}K^{\beta\ast}P^T(\omega_1+\omega_2)+F) + \frac{1}{2}S_1P_{\beta\beta}$$
and
$$B_1(\hat{\phi})= \frac{1}{2}T_2(2\tilde{Z}K^{\theta\ast}P^T(\omega_1+\omega_2)+G) + \frac{1}{2}S_2P_{\theta\theta}.$$

Further, let $\check{\beta}$ and $\check{\phi}$ be the estimators obtained as the root of the modified score function (\ref{modscore}). Then
we also obtain the following formulae for the ${\cal O}(n^{-1})$ biases of $\hat{\mu}$ and $\hat{\phi}$, respectively:
$$B_2(\hat{\mu}) = \frac{1}{2}T_1(2\tilde{X}(\hat{\beta}-\check{\beta})+F) + \frac{1}{2} S_1P_{\beta\beta}$$
and
$$B_2(\hat{\phi}) = \frac{1}{2}T_2(2\tilde{Z}(\hat{\theta}-\check{\theta})+G) + \frac{1}{2} S_2P_{\theta\theta}.$$

Lastly, we can use the bootstrap-based ${\cal O}(n^{-1})$ biases to define, bias corrected estimators of $\hat{\mu}$ and $\hat{\phi}$ to this order.
Then, let $\hat{B}_{F_{\hat{\zeta}}}(\hat{\beta})$ be the vector formed by the first $k$ elements of the vector $\hat{B}_{F_{\hat{\zeta}}}(\hat{\zeta}, \zeta)$
defined in equation (\ref{biasboot}), $\hat{B}_{F_{\hat{\zeta}}}(\hat{\theta})$ be the vector formed by the last $h$ elements of the vector $\hat{B}_{F_{\hat{\zeta}}}(\hat{\zeta}, \zeta)$,
and define $\hat{B}_{\hat{F}}(\hat{\beta})$ and $\hat{B}_{\hat{F}}(\hat{\theta})$ analogously from the vector $\hat{B}_{\hat{F}}(\hat{\zeta}, \zeta)$ also in equation (\ref{biasboot}).
Thus, we have the following alternative expressions for the ${\cal O}(n^{-1})$ biases of $\hat{\mu}$ and $\hat{\phi}$, respectively:
$$B_3(\hat{\mu}) = \frac{1}{2}T_1(2\tilde{X}\hat{B}_{F_{\hat{\zeta}}}(\hat{\beta})+F) + \frac{1}{2} S_1 P_{\beta\beta}\hbox{~and~}B_4(\hat{\mu}) = \frac{1}{2}T_1 (2\tilde{X}\hat{B}_{\hat{F}}(\hat{\beta})+F) + \frac{1}{2} S_1 P_{\beta\beta},$$
and
$$B_3(\hat{\phi}) = \frac{1}{2}T_2(2\tilde{Z}\hat{B}_{F_{\hat{\zeta}}}(\hat{\theta}) + G) + \frac{1}{2} S_2 P_{\theta\theta} \hbox{~and~}B_4(\hat{\phi}) = \frac{1}{2}T_2 (2\tilde{Z}\hat{B}_{\hat{F}}(\hat{\theta})+G) + \frac{1}{2} S_2 P_{\theta\theta}.$$

Therefore, we are now able to define the following second-order bias-corrected estimators for $\hat{\mu}$ and $\hat{\phi}$:
$$\tilde{\mu} = \hat{\mu} - \hat{B}_1(\hat{\mu}),\quad \check{\mu} = \hat{\mu} - \hat{B}_2(\hat{\mu}),\quad\overline{\mu}_1 = \hat{\mu}-\hat{B}_3(\hat{\mu})\hbox{~~and~~} \overline{\mu}_2 = \hat{\mu}-\hat{B}_4(\hat{\mu})$$
and
$$\tilde{\phi} = \hat{\phi} - \hat{B}_1(\hat{\phi}),\quad \check{\phi} = \hat{\phi} - \hat{B}_2(\hat{\phi}),\quad\overline{\phi}_1 = \hat{\phi}-\hat{B}_3(\hat{\phi})\hbox{~~and~~} \overline{\phi}_2 = \hat{\phi}-\hat{B}_4(\hat{\phi}),$$
where, for $j=1,2,3$ and $4$, $\hat{B}_j(\cdot)$ denotes the MLE of $B_j(\cdot)$, that is, the unknown parameters are replaced by their MLEs. Finally, one can also use bootstrap techniques to obtain directly the biases of $\hat{\mu}$ and $\hat{\phi}$, therefore, denote by $\hat{B}_{F_{\hat{\mu},\hat{\phi}}}(\hat{\mu})$ the vector of ${\cal O}(n^{-1})$ parametric bootstrap biases for $\hat{\mu}$, and by $\hat{B}_{F_{\hat{\mu},\hat{\phi}}}(\hat{\phi})$ the vector of ${\cal O}(n^{-1})$ parametric bootstrap biases for $\hat{\phi}$ defined in equation (\ref{biasboot}). Further, define $\hat{B}_{\hat{F}}(\hat{\mu})$ and $\hat{B}_{\hat{F}}(\hat{\phi})$ analogously from the vector $\hat{B}_{\hat{F}}(\hat{\zeta}, \zeta)$ also in equation (\ref{biasboot}). Hence, we can also define the following bias-corrected estimators of $\hat{\mu}$ and $\hat{\phi}$:
$$\overline{\overline{\mu}}_1 = \hat{\mu} - \hat{B}_{F_{\hat{\mu},\hat{\phi}}}(\hat{\mu}) \hbox{~and~}
\overline{\overline{\mu}}_2 = \hat{\mu} - \hat{B}_{\hat{F}}(\hat{\mu}),$$
and
$$\overline{\overline{\phi}}_1 = \hat{\phi} - \hat{B}_{F_{\hat{\mu},\hat{\phi}}}(\hat{\phi}) \hbox{~and~}
\overline{\overline{\phi}}_2 = \hat{\phi} - \hat{B}_{\hat{F}}(\hat{\phi}).$$

We give in Tables \ref{tabelalink1} and \ref{tabelalink2} the most common link functions for $g_1$ and $g_2$,
respectively, together with their first and second derivatives. We believe this will help the practioners
that may be interested in applying our results. For Table \ref{tabelalink1} $\Phi(\cdot)$ denotes
the standard normal distribution function, $f(x) = 1/\sqrt{2\pi}\exp\{-1/2 x^2\}$ is the density of a
standard normal distribution and $f'(x) = -x/\sqrt{2\pi} \exp\{-1/2 x^2\}$ is the derivative of the density
of a standard normal distribution.

\begin{table}[htb]
\caption{Values of $d\mu/d\eta_1$, $d^2\mu/d\eta_1^2$ for the most common link functions.}
\begin{center}
\begin{tabular}{ccc}
\hline
Link function & Formula & $d\mu/d\eta_1$\\
\hline
Logit & $\log(\mu/(1-\mu)) = \eta_1$ & $\mu(1-\mu)$\\
Probit& $\Phi^{-1}(\mu)=\eta_1$ & $f(\Phi^{-1}(\mu))$\\
Comp. Log-Log& $\log(-\log(1-\mu)) = \eta_1$ & $-\log(1-\mu)/(1-\mu)$\\
\hline
\hline
Link function & Formula & $d^2\mu/d\eta_1^2$\\
\hline
Logit & $\log(\mu/(1-\mu)) = \eta_1$ &  $\mu(1-\mu)(1-2\mu)$\\
Probit& $\Phi^{-1}(\mu)=\eta_1$ & $f'(\Phi^{-1}(\mu))$\\
Comp. Log-Log& $\log(-\log(1-\mu)) = \eta_1$ & $-(1-\mu)\log(1-\mu)(1+\log(1-\mu))$\\
\hline
\end{tabular}
\end{center}
\label{tabelalink1}
\end{table}
\begin{table}[htb]
\caption{Values of $d\phi/d\eta_2$, $d^2\phi/d\eta_2^2$ for the most common link functions.}
\begin{center}
\begin{tabular}{cccc}
\hline
Link function & Formula & $d\phi/d\eta_2$ & $d^2\phi/d\eta_2^2$\\
\hline
Identity & $\phi = \eta_2$ & $1$ & $0$\\
Log& $\log(\phi)=\eta_2$ & $\phi$ & $\phi$\\
Square root& $\sqrt{\phi} = \eta_2$ & $2\phi$ & $2$\\
\hline
\end{tabular}
\end{center}
\label{tabelalink2}
\end{table}

\section{Some special cases}\label{partcases}
In this section we consider some special models that commonly arise in the practical use, namely, the linear beta regression model, the linear beta
regression model with dispersion covariates, the nonlinear beta regression model and the nonlinear beta regression model with linear dispersion covariates.
Further, we study these models in full detail, that is, we give closed form expressions for their score vector, Fisher's information matrix and for the
${\cal O}(n^{-1})$ biases of the MLEs of $\beta$, $\theta$, $\mu$ and $\phi$.
\subsection{The linear beta regression model}
For linear beta regression models we have, in (\ref{regr}), $g_1(\mu_i) = g(\mu_i)$, where $g(\cdot)$ is some link function, $g_2(\phi_i) = \phi_i$, and
further, we can write (\ref{regr}) as
$$g(\mu_i) = \eta_i = x_i^T \beta\hbox{~~and~~}\phi_i = \phi,$$
where $\phi>0$ is a constant, i.e., we have that in this case $\tilde{X} = X$ and $\tilde{Z} = \mathbf{1}$, where $X$ is the matrix of covariates with rows given by $x_i^T$, and
the parameters $\beta\in\mathbb{R}^k$ and $\phi\in (0,\infty)$.
Furthermore, the score vector (\ref{scorevec}) becomes
\begin{equation}\label{scorelinear}
\begin{array}{ccl}
U_\beta(\beta,\phi) &=& \phi X^T T (y^\ast - \mu^\ast),\\
U_\phi(\beta,\phi) &=& \sum_{i=1}^n v_i,
\end{array}
\end{equation}
where $T = {\rm diag}(d\mu_i/d\eta_i)$ and $y^\ast,\mu^\ast$, and $v_i$ are defined in page \pageref{scorevec}. Moreover, the matrices $P$ and $W$ defined by
equations (\ref{matrizp}) and (\ref{matrizw}) becomes, respectively,
$$P = \left(\begin{array}{cc}
X&0\\
0&\mathbf{1}
\end{array}\right),$$
and
\begin{equation}\label{matrizwlinear}
W = \left(\begin{array}{cc}
W_{\beta\beta}&W_{\beta\phi}\\
W_{\beta\phi} & W_{\phi\phi},
\end{array}\right),
\end{equation}
with
$$W_{\beta\beta} = {\rm diag} \left(\phi^2 a_i \left(\frac{d\mu_i}{d\eta_i}\right)^2\right),$$
$$W_{\beta\phi} = {\rm diag} \left(\phi \{\mu_i a_i -\psi'((1-\mu_i)\phi)\} \left(\frac{d\mu_i}{d\eta_i}\right)\right),$$
$$W_{\phi\phi} = {\rm diag}(b_i).$$
Here, $a_i$ and $b_i$ are as defined in page \pageref{scorevec}. Therefore, we have the following expression for the Fisher's information matrix
for the parameter vector $\zeta = (\beta^T,\phi)^T$
\begin{equation}\label{fisherlinear}
K(\zeta) = P^TWP.
\end{equation}
By means of simple calculations it is possible to conclude that equations (\ref{scorelinear}) and (\ref{fisherlinear}) agree with the expressions for
the score vector and Fisher's information matrix given in Ferrari and Cribari-Neto (2004).

We now move to the bias correction. Note initially that $\xi_2$ given in (\ref{biaszeta}) vanishes since, for this model, both $\eta_1$ and $\eta_2$ are linear functions
of $\beta$ and $\theta$, respectively. Thus, if $P$ and $W$ are the matrices defined for this model, equation (\ref{biaszeta}) equals simply
$$B(\hat{\zeta}) = (P^TWP)^{-1}P^TW\xi_1,$$
which is easy to see that agrees with the finds of Ospina et al. (2006). Further, $\omega_2$ in equation (\ref{modscore}) also vanishes which gives us
the expression
$$U^\ast(\zeta) = U(\zeta) + P^T\omega_1,$$
and it is also easy to see that it agrees with the finds of Ospina et al. (2006). Finally, following equation (\ref{biasmu}), the second order bias of $\hat{\mu}$ can be expressed, in
matrix notation, as
$$B(\hat{\mu}) = TXB(\hat{\beta}) + \frac{1}{2}S_1P_{\beta\beta},$$
with $T$ defined in the begining of this subsection and $S_1$ was defined in page \pageref{biasmu}.
\subsection{The linear beta regression model with dispersion covariates}

This class of models generalizes the linear beta regression models considered in the last subsection by letting the precision parameter $\phi$ vary
through a linear regression structure. More precisely, for this model the equation (\ref{regr}) becomes
$$g_1(\mu_i) = \eta_{1i} = x_i^T\beta\hbox{~~and~~}g_2(\phi_i) = \eta_{2i} = z_i^T\theta,$$
where $\beta\in\mathbb{R}^k$ and $\theta\in\mathbb{R}^h$. Then, we have that for this model $\tilde{X} = X$ and $\tilde{Z} = Z$, where $X$ is the matrix of covariates with rows given by $x_i^T$
and $Z$ is the matrix of covariates with rows given by $z_i^T$. The score vector for this model is identical to the one given in expression (\ref{scorevec}),
only that in this case $\tilde{X}$ and $\tilde{Z}$ must be replaced by $X$ and $Z$, respectively.

Now, the matrix $P$ defined in (\ref{matrizp}) becomes for this model
$$P = \left(\begin{array}{cc}
X&0\\
0&Z
\end{array}\right),$$
further, let $W$ be the matriz defined in (\ref{matrizw}), thus, we have the following expression for the Fisher's information matrix
for the parameter vector $\zeta = (\beta^T,\theta)^T$
$$K(\zeta) = P^TWP.$$

Moving to bias correction, since both $\eta_1$ and $\eta_2$ are linear functions of $\beta$ and $\theta$, respectively, we have that $\xi_2$ in equation (\ref{biaszeta}) vanishes
and thus the ${\cal O}(n^{-1})$ bias of the MLEs of the parameter vector $\zeta$ is
$$B(\hat{\zeta}) = (P^TWP)^{-1}P^TW\xi_1.$$
Further, since $\omega_2$ in (\ref{modscore}) vanishes,  the modified score to obtain the estimator through the preventive method is given by
$$U^\ast(\zeta) = U(\zeta) + P^T\omega_1,$$
and, finally, using equations (\ref{biasmu}) and (\ref{biasphi}), the second order biases of $\hat{\mu}$ and $\hat{\phi}$ are given by
$$B(\hat{\mu}) = T_1XB(\hat{\beta}) + \frac{1}{2}S_1P_{\beta\beta},$$
and
$$B(\hat{\phi}) = T_2ZB(\hat{\theta}) + \frac{1}{2} S_2 P_{\theta\theta},$$
where $T_1$ and $T_2$ were defined in page \pageref{scorevec}, $S_1$ and $S_2$ were defined in page \pageref{biasmu}, $P_{\beta\beta}$ and $P_{\theta\theta}$
were defined in page \pageref{formulas}.
\subsection{The nonlinear beta regression model}
We now consider the nonlinear beta regression models. For this models, the expressions in (\ref{regr}) turns to be
$g_1(\mu_i) = g(\mu_i)$, where $g(\cdot)$ is some link function, $g_2(\phi_i) = \phi_i$, and
further, we can write (\ref{regr}) as
$$g(\mu_i) = \eta_i = f(x_i^T; \beta)\hbox{~~and~~}\phi_i = \phi,$$
where $\phi>0$ is a constant, i.e., and thus, in this case $\tilde{X}$ remains the same and $\tilde{Z} = \mathbf{1}$,
the parameters $\beta\in\mathbb{R}^k$ and $\phi\in (0,\infty)$.
Furthermore, the score vector (\ref{scorevec}) becomes
\begin{equation*}
\begin{array}{ccl}
U_\beta(\beta,\phi) &=& \phi \tilde{X}^T T (y^\ast - \mu^\ast),\\
U_\phi(\beta,\phi) &=& \sum_{i=1}^n v_i,
\end{array}
\end{equation*}
where $T$ is as defined in equation (\ref{scorelinear}), and $v_i$ are defined in page \pageref{scorevec}. Moreover, the matrix $P$ defined by
equation (\ref{matrizp}), becomes
$$P = \left(\begin{array}{cc}
\tilde{X}&0\\
0&\mathbf{1}
\end{array}\right),$$
and the matrix $W$ defined by equation (\ref{matrizw}) is actually the same as the matrix given in equation (\ref{matrizwlinear}). Therefore, the Fisher's information matrix
for the parameter vector $\zeta = (\beta^T,\phi)^T$ can also be written as $K(\zeta) = P^TWP$.

We now turn to bias correction. For this model, the vector $\xi_2$ in equation (\ref{biaszeta}) can be written as $\xi_2 = W^{-1}\tilde{\omega}_2$,
where $\tilde{\omega}_2$ is given below:
\begin{equation}\label{omegamodificado}
\tilde{\omega}_2 = \left(\begin{array}{c}
- N_1 F\\
N_2 F
\end{array}\right),
\end{equation}
where $N_1$ and $N_2$ are given in equations (\ref{matrizn1}) and (\ref{matrizn2}) of Appendix, and $F$ is given in page \pageref{formulas}. Thus, using $P, W$ and $\xi_2$ defined
in this Subsection, the ${\cal O}(n^{-1})$ bias for the MLEs of the parameter vector $\zeta$ is given by
$$B(\hat{\zeta}) = (P^TWP)^{-1}P^TW(\xi_1+\xi_2),$$
$\xi_1$ being as defined in formula (\ref{biaszeta}). Further, the modified score to obtain the estimator through the preventive method is given by
$$U^\ast(\zeta) = U(\zeta) + P^T(\omega_1+\tilde{\omega}_2).$$
Moreover, the second order bias of $\hat{\mu}$ can be written, following equation (\ref{biasmu}), as
$$B(\hat{\mu}) =  \frac{1}{2}T (2\tilde{X}B(\hat{\beta})+F) + \frac{1}{2} S_1 P_{\beta\beta},$$
with $S_1$ as defined in page \pageref{biasmu}.
\subsection{The nonlinear beta regression model with linear dispersion covariates}

This class of models generalizes the nonlinear beta regression models considered in the last subsection by letting the precision parameter $\phi$ vary
through a linear regression structure, such as the linear beta regression with dispersion covariates generalizes the linear beta regression.
More precisely, for this model the equation (\ref{regr}) becomes
$$g_1(\mu_i) = \eta_{1i} = f(x_i^T;\beta)\hbox{~~and~~}g_2(\phi_i) = \eta_{2i} = z_i^T\theta,$$
where $\beta\in\mathbb{R}^k$ and $\theta\in\mathbb{R}^h$. Then, we have that for this model $\tilde{X}$ remaining the same, and $\tilde{Z} = Z$, where
$Z$ is the matrix of covariates with rows given by $z_i^T$. The score vector for this model is identical to the one given in expression (\ref{scorevec}),
only that, in this case $\tilde{Z}$ must be replaced by $Z$.

Now, the matrix $P$ defined in (\ref{matrizp}) becomes for this model
$$P = \left(\begin{array}{cc}
\tilde{X}&0\\
0&Z
\end{array}\right),$$
further, let $W$ be the matriz defined in (\ref{matrizw}), thus, we have the following expression for the Fisher's information matrix
for the parameter vector $\zeta = (\beta^T,\theta)^T$
$$K(\zeta) = P^TWP.$$

Moving to bias correction, for this model the vector $\xi_2$ in equation (\ref{biaszeta}) can be written as $\xi_2 = W^{-1}\tilde{\omega}_2$,
where $\tilde{\omega}_2$ was defined in equation (\ref{omegamodificado}). Thus, using $P$ and $\xi_2$ defined
in this Subsection, the ${\cal O}(n^{-1})$ bias for the MLEs of the parameter vector $\zeta$ is given by
$$B(\hat{\zeta}) = (P^TWP)^{-1}P^TW(\xi_1+\xi_2),$$
$\xi_1$ being as defined in formula (\ref{biaszeta}). Further, the modified score to obtain the estimator through the preventive method is given by
$$U^\ast(\zeta) = U(\zeta) + P^T(\omega_1+\tilde{\omega}_2).$$
Moreover, the second order biases of $\hat{\mu}$ and $\hat{\phi}$ can be written, following equation (\ref{biasmu}), respectively, as
$$B(\hat{\mu}) =  \frac{1}{2}T_1 (2\tilde{X}B(\hat{\beta})+F) + \frac{1}{2} S_1 P_{\beta\beta},$$
and
$$B(\hat{\phi}) = T_2ZB(\hat{\theta}) + \frac{1}{2} S_2 P_{\theta\theta},$$
with $T_1$ and $T_2$ as defined in page \pageref{scorevec}, and, $S_1$ and $S_2$ as defined in page \pageref{biasmu}.
\section{Numerical results}
In this section we present the results of some Monte Carlo simulation experiments,
where we study the finite-sample distributions of the MLEs of $\beta$ and $\theta$ along with their corrected versions proposed in this paper. The first experiment uses a
logit link in a nonlinear model for the regression parameters and a log link in a nonlinear model for the precision
parameter
$${\rm logit}\mu_i = \beta_0 + {\beta_1}x_{1,i} +  x_{2,i}^{\beta_2},$$
$${\rm log}\phi_i = \theta_0 + \theta_1 x_{1,i} + x_{2,i}^{\theta_2},\quad i=1,\ldots,n,$$ 
where the true values of the parameters were taken as $\beta_0 = 1.5$, $\beta_1 = 0.5$ and $\beta_2 = 2$;
and $\theta_0 = 1.7, \theta_1 = 0.7$ and $\theta_2 = 3$. Note also that here
the elements of the $n\times 3$ matrix $\tilde{X}$ are: $\tilde{X}(\beta)_{i,1} = 1; \tilde{X}(\beta)_{i,2} = x_{1,i}$, and
$\tilde{X}(\beta)_{i,3} = \log(x_{2,i}) x_{2,i}^{\beta_2}$. The explanatory variables $x_1$ and $x_2$ were generated
from the standard normal and uniform U$(1,2)$ distributions, respectively, for $n=20,40$ and $60$.
The values of $x_1$ and $x_2$ were held constant throughout the simulations.
The total number of Monte Carlo replications was set at $5,000$ for each sample size.
All simulations were performed using the software {\tt Ox}.
\begin{table}[t!]
\caption{Simulation results for $n=20$.}
\begin{center}
\begin{tabular}{lccccc}
\hline
Parameter & MLE& Cox-Snell & Firth & p-boot & np-boot\\
\hline
$\beta_0$  &1.5147 &1.5032 &1.5253 &1.5019 &1.5061 \\
Bias       &0.0147 &0.0032 &0.0253 &0.0019 &0.0061\\ 
Variance   &0.0628 &0.0627 &0.0667 &0.0627 &0.0677\\
MSE        &0.0630 &0.0627 &0.0674 &0.0627 &0.0677\\\\
$\beta_1$  &0.5001 &0.5001 &0.5002 &0.5005 &0.5017         \\
Bias       &0.0001 &0.0001 &0.0002 &0.0005 &0.0017\\
Variance   &0.0126 &0.0126 &0.0132 &0.0125 &0.0143\\
MSE        &0.0126 &0.0126 &0.0132 &0.0125 &0.0143\\\\
$\beta_2$  &1.9905 &1.9968 &1.9847 &1.9982 &2.0036\\
Bias       &-0.0095&-0.0032&-0.0153&-0.0018&0.0036\\
Variance   &0.0119 &0.0118 &0.0129 &0.0117 &0.0125\\
MSE        &0.0120 &0.0118 &0.0130 &0.0117 &0.0125\\\\
$\theta_0$ &1.9512 &1.7406 &2.1388 &1.7045 & 1.8107\\
Bias       &0.2512 &0.0406 &0.4388 &0.0045 & 0.1107\\
Variance   &0.5615 &0.5100 &0.6248 &0.4833 &0.4739\\
MSE        &0.6247 &0.5116 &0.8173 &0.4833 &0.4861\\\\
$\theta_1$ &0.7091 &0.6928 &0.7331 &0.6842 &0.7060\\
Bias       &0.0091 &-0.0072&0.0331 &-0.1584&0.0060\\
Variance   &0.2881 &0.2481 &0.3651 &0.2231 &0.2369\\
MSE        &0.2882 &0.2481 &0.3662 &0.2234 &0.2369\\\\
$\theta_2$ &3.0261 &3.0061 &3.0440 &3.0043 &3.0757\\
Bias       &0.0261 &0.0061 &0.0440 &0.0043 &0.0757\\
Variance   &0.0706 &0.0610 &0.0805 &0.0514 &0.0471\\
MSE        &0.0720 &0.0611 &0.0825 &0.0514 &0.0528\\
\hline
\end{tabular}
\end{center}
\label{resulsimul}
\end{table}

\begin{sidewaystable}
\caption{Estimated values of $\mu$ for $n=20$.}
\begin{center}
\begin{tabular}{cccccccccccccccc}
\hline
$i$ & \multicolumn{2}{c}{MLE}& \multicolumn{2}{c}{Cox-Snell}  & \multicolumn{2}{c}{p-boot} &\multicolumn{2}{c}{p-boot} & \multicolumn{2}{c}{np-boot}& \multicolumn{2}{c}{np-boot}\\
     & $\hat{\mu}_i$        &MSE$\times 10^5$&  $\tilde{\mu}_i$        &MSE$\times 10^5$& $\overline{\mu}_{1,i}$        &MSE$\times 10^5$& $\overline{\overline{\mu}}_{1,i}$        &MSE$\times 10^5$& $\overline{\mu}_{2,i}$        &MSE$\times 10^5$& $\overline{\overline{\mu}}_{2,i}$        &MSE$\times 10^5$\\
    \hline
 $1$&0.99576&0.0108&0.99576&0.0108&0.99577&0.0107&0.99577&0.0107&0.99581&0.0128&0.99585&0.0122 \\
 $2$&0.99355&0.2276&0.99361&0.2221&0.99364&0.2203&0.99361&0.2223&0.99401&0.2616&0.99369&0.2571 \\
 $3$&0.95051&0.0001&0.95069&1.0474&0.95081&1.0440&0.95065&1.0492&0.95252&1.1269&0.95092&1.1430 \\
 $4$&0.98047&0.7630&0.98049&0.7620&0.98051&0.7594&0.98050&0.7616&0.98084&0.8367&0.98070&0.8373 \\
 $5$&0.95597&7.5592&0.95612&7.5248&0.95621&7.4940&0.95609&7.5313&0.95755&8.1211&0.95635&8.2285 \\
 $6$&0.99447&0.0152&0.99447&0.0152&0.99447&0.0152&0.99448&0.0151&0.99451&0.0168&0.99458&0.0169 \\
 $7$&0.97837&1.1784&0.97838&1.1761&0.97842&1.1737&0.97838&1.1783&0.97897&1.2768&0.97858&1.2786 \\
 $8$&0.99383&0.0126&0.99383&0.0126&0.99383&0.0126&0.99384&0.0126&0.99387&0.0143&0.99394&0.0146 \\
 $9$&0.99537&0.0117&0.99537&0.0117&0.99537&0.0117&0.99538&0.0116&0.99542&0.0133&0.99547&0.0129 \\
$10$&0.98111&1.1470&0.98116&1.1389&0.98120&1.1357&0.98116&1.1418&0.98183&1.2585&0.98133&1.2525 \\
$11$&0.99451&0.0115&0.99450&0.0115&0.99450&0.0115&0.99451&0.0114&0.99455&0.0134&0.99460&0.0134 \\
$12$&0.99002&0.1023&0.99002&0.1022&0.99003&0.1020&0.99004&0.1020&0.99013&0.1126&0.99018&0.1133 \\
$13$&0.97723&2.1492&0.97732&2.1264&0.97739&2.1185&0.97732&2.1327&0.97832&2.3582&0.97750&2.3507 \\
$14$&0.99222&0.1662&0.99224&0.1646&0.99226&0.1640&0.99225&0.1648&0.99251&0.1888&0.99235&0.1856 \\
$15$&0.99169&0.0522&0.99167&0.0523&0.99168&0.0523&0.99168&0.0523&0.99178&0.0590&0.99180&0.0586 \\
$16$&0.96273&4.7219&0.96282&4.7072&0.96289&4.6905&0.96280&4.7116&0.96393&5.0719&0.96305&5.1305 \\
$17$&0.97674&2.0972&0.97687&2.0702&0.97693&2.0568&0.97688&2.0675&0.97758&2.2964&0.97711&2.3020 \\
$18$&0.95009&1.3879&0.95039&1.3752&0.95056&1.3703&0.95035&1.3794&0.95288&1.5040&0.95061&1.5142 \\
$19$&0.94657&1.5431&0.94695&1.5243&0.94712&1.5146&0.94691&1.5251&0.94909&1.6708&0.94717&1.6917 \\
$20$&0.92738&3.6923&0.92811&3.6351&0.92840&3.6103&0.92804&3.6378&0.93157&4.0065&0.92828&4.0608 \\
\hline
\end{tabular}
\end{center}
\label{tabelasimulmu1}
\end{sidewaystable}

\begin{sidewaystable}
\caption{Estimated values of $\phi$ for $n=20$.}
\begin{center}
\begin{tabular}{ccccccccccccccccc}
\hline
$i$ &True & \multicolumn{2}{c}{MLE}& \multicolumn{2}{c}{Cox-Snell}  & \multicolumn{2}{c}{p-boot} & \multicolumn{2}{c}{np-boot}\\
&$\phi_i$ &  $\hat{\phi}_i$       &MSE$\times 10^{-5}$&  $\tilde{\phi}$        &MSE$\times 10^{-5}$& $\overline{\overline{\phi}}_{1,i}$        &MSE$\times 10^{-5}$& $\overline{\overline{\phi}}_{2,i}$       &MSE$\times 10^{-5}$\\
    \hline
 $1$&12218 &32123 &48311 &12500 &3516.6&8542.3&831.41&13848 &21445  \\
 $2$&1512.2&4977.5&6716.2&1567.4&212.00&873.67&97.962&1991.9&809.20 \\
 $3$&33.710&55.534&0.0308&35.446&0.0082&33.122&0.0071&39.450&0.0138 \\
 $4$&381.74&652.63&3.5711&404.02&0.8616&374.10&0.6805&523.56&2.3280 \\
 $5$&49.260&81.864&0.0911&52.560&0.0236&49.382&0.0187&60.855&0.0544 \\
 $6$&8602.5&22211 &19727 &8858.7&1522.2&6182.9&308.90&9874.8&9677.1 \\
 $7$&168.67&267.63&0.4576&174.93&0.1307&163.20&0.1128&218.12&0.2947 \\
 $8$&5343.4&11400 &2349.0&5613.7&34.533&4558.3&121.79&7036.3&843.28 \\
 $9$&12920 &37410 &90373 &12865 &4081.7&7660.0&2134.6&12609 &58969  \\
$10$&191.66&343.11&2.0051&203.41&0.4408&182.27&0.2683&248.99&0.7876 \\
$11$&6293.7&13636 &3631.7&6597.4&0.0518&5301.0&180.91&8269.6&1213.7 \\
$12$&2235.6&4604.2&304.53&2362.9&50.021&2000.5&23.615&3035.0&124.77 \\
$13$&122.77&237.92&1.5913&132.08&0.2858&114.87&0.1369&152.74&0.4410 \\
$14$&1288.6&2744.5&306.24&1377.5&44.117&1142.9&15.643&1843.3&82.982 \\
$15$&1706.1&2935.0&55.657&1767.5&13.164&1589.4&9.5611&2382.9&32.945 \\
$16$&66.492&106.15&0.0928&69.929&0.0027&66.068&0.0234&83.020&0.0559 \\
$17$&375.66&876.02&33.651&412.62&3.7010&351.16&3.1617&520.18&23.952 \\
$18$&28.969&57.292&0.0949&30.760&0.0153&26.626&0.0068&30.897&0.0227 \\
$19$&43.286&98.112&1.9018&50.435&0.1725&44.321&0.1744&57.730&2.6463 \\
$20$&24.754&81.680&9.3312&30.351&0.1859&22.880&0.1946&39.086&15.773 \\
\hline
\end{tabular}
\end{center}
\label{tabelasimulphi1}
\end{sidewaystable}

\begin{table}[t!]
\caption{Simulation results for $n=40$.}
\begin{center}
\begin{tabular}{lccccc}
\hline
Parameter & MLE& Cox-Snell & Firth & p-boot & np-boot\\
\hline
$\beta_0$&1.5039   &1.4991   &1.5079   &1.4962   &1.4946   \\
$\beta_0$   &1.5039   &1.4991   &1.5079   &1.4962   &1.4946\\
Bias        &0.0039   &-0.0009  &0.0079   &-0.0038  &-0.0054\\ 
Variance    &0.0329   &0.0330   &0.0337   &0.0329   &0.0339\\
MSE         &0.0329   &0.0330   &0.0338   &0.0329   &0.0339\\\\
$\beta_1$   &0.5029   &0.5028   &0.5033   &0.5030   &0.5037\\
Bias        &0.0029   &0.0028   &0.0033   &0.0030   &0.0037\\
Variance    &0.0077   &0.0077   &0.0079   &0.0077   &0.0084\\
MSE         &0.0077   &0.0077   &0.0079   &0.0077   &0.0084\\\\
$\beta_2$   &1.9960   &1.9993   &1.9930   &2.0014   &2.0069\\
Bias        &-0.0040  &-0.0007  &-0.0070  &0.0014   &0.0069\\
Variance    &0.0146   &0.0145   &0.0151   &0.0144   &0.0150\\
MSE         &0.0146   &0.0145   &0.0151   &0.0144   &0.0150\\\\
$\theta_0$  &1.7460   &1.6884   &1.7865   &1.6972   &1.7033\\
Bias        &0.0460   &-0.0116  &0.0865   &-0.0028  &0.0033   \\
Variance    &0.1895   &0.1755   &0.2108   &0.1708   &0.1811  \\
MSE         &0.1916   &0.1757   &0.2183   &0.1708   &0.1811  \\\\
$\theta_1$  &0.7264   &0.7110   &0.7480   &0.7043   &0.7021  \\
Bias        &0.0264   &0.0110   &0.0480   &0.0043   &0.0021  \\
Variance    &0.0744   &0.0696   &0.0816   &0.0697   &0.0727 \\
MSE         &0.0751   &0.0698   &0.0839   &0.0697   &0.0727 \\\\
$\theta_2$  &3.0701   &3.0220   &3.1226   &3.0067   &3.0632   \\
Bias        &0.0701   &0.0220   &0.1226   &0.0067   &0.0632   \\
Variance    &0.0709   &0.0652   &0.0791   &0.0597   &0.0633 \\
MSE         &0.0758   &0.0657   &0.0941   &0.0597   &0.0673 \\
\hline
\end{tabular}
\end{center}
\label{resulsimul2}
\end{table}

\begin{sidewaystable}
\caption{Estimated values of $\mu$ for $n=40$ for observations $1$ until $20$.}
\begin{center}
\begin{tabular}{cccccccccccccccc}
\hline
$i$ & \multicolumn{2}{c}{MLE}& \multicolumn{2}{c}{Cox-Snell}  & \multicolumn{2}{c}{p-boot} &\multicolumn{2}{c}{p-boot} & \multicolumn{2}{c}{np-boot}& \multicolumn{2}{c}{np-boot}\\
     & $\hat{\mu}_i$        &MSE$\times 10^5$&  $\tilde{\mu}_i$        &MSE$\times 10^5$& $\overline{\mu}_{1,i}$        &MSE$\times 10^5$& $\overline{\overline{\mu}}_{1,i}$        &MSE$\times 10^5$& $\overline{\mu}_{2,i}$        &MSE$\times 10^5$& $\overline{\overline{\mu}}_{2,i}$        &MSE$\times 10^5$\\
    \hline
 $1$&0.99163&0.0499&0.99162&0.0501&0.99162&0.0501&0.99163&0.0500&0.99167&0.0527&0.99170&0.0534 \\
 $2$&0.99236&0.1183&0.99238&0.1176&0.99239&0.1176&0.99238&0.1177&0.99246&0.1267&0.99242&0.1264 \\
 $3$&0.99225&0.0955&0.99226&0.0953&0.99226&0.0953&0.99227&0.0950&0.99232&0.1004&0.99234&0.1001 \\
 $4$&0.95825&3.3785&0.95839&3.3684&0.95836&3.3660&0.95833&3.3704&0.95854&3.5272&0.95841&3.5644 \\
 $5$&0.97837&0.5878&0.97839&0.5886&0.97838&0.5881&0.97838&0.5880&0.97846&0.6187&0.97849&0.6220 \\
 $6$&0.97923&0.3741&0.97923&0.3759&0.97921&0.3757&0.97921&0.3757&0.97927&0.3893&0.97931&0.3922 \\
 $7$&0.98064&0.2667&0.98062&0.2679&0.98060&0.2682&0.98061&0.2683&0.98065&0.2717&0.98068&0.2739 \\
 $8$&0.97723&0.4311&0.97722&0.4333&0.97720&0.4333&0.97720&0.4334&0.97725&0.4431&0.97729&0.4463 \\
 $9$&0.98934&0.1064&0.98933&0.1067&0.98933&0.1067&0.98934&0.1065&0.98938&0.1120&0.98942&0.1128 \\
$10$&0.96028&3.9369&0.96049&3.9106&0.96045&3.9195&0.96040&3.9269&0.96064&4.0261&0.96040&4.0708 \\
$11$&0.98530&0.1122&0.98528&0.1129&0.98527&0.1129&0.98527&0.1129&0.98531&0.1154&0.98536&0.1173 \\
$12$&0.95741&3.4003&0.95755&3.3922&0.95751&3.3906&0.95749&3.3950&0.95769&3.5413&0.95755&3.5779 \\
$13$&0.96244&3.7194&0.96266&3.6900&0.96261&3.6983&0.96257&3.7058&0.96282&3.8067&0.96257&3.8523 \\
$14$&0.99558&0.0243&0.99558&0.0240&0.99558&0.0243&0.99559&0.0243&0.99563&0.0265&0.99563&0.0263 \\
$15$&0.98461&0.1285&0.98459&0.1291&0.98457&0.1293&0.98458&0.1293&0.98461&0.1311&0.98465&0.1329 \\
$16$&0.91254&2.1237&0.91303&2.1175&0.91290&2.1211&0.91280&2.1242&0.91332&2.1894&0.91266&2.2149 \\
$17$&0.93795&1.3121&0.93844&1.2955&0.93842&1.2935&0.93835&1.2965&0.93892&1.3725&0.93839&1.3964 \\
$18$&0.98201&0.3038&0.98201&0.3043&0.98200&0.3048&0.98199&0.3051&0.98205&0.3117&0.98206&0.3135 \\
$19$&0.89786&3.6970&0.89866&3.6690&0.89856&3.6697&0.89841&3.6768&0.89930&3.8638&0.89826&3.9294 \\
$20$&0.90462&3.6712&0.90552&3.6270&0.90547&3.6245&0.90532&3.6330&0.90631&3.8445&0.90522&3.9185 \\
\hline
\end{tabular}
\end{center}
\label{tabelasimulmu21}
\end{sidewaystable}
\begin{sidewaystable}
\caption{Estimated values of $\mu$ for $n=40$ for observations $21$ until $40$.}
\begin{center}
\begin{tabular}{cccccccccccccccc}
\hline
$i$ & \multicolumn{2}{c}{MLE}& \multicolumn{2}{c}{Cox-Snell}  & \multicolumn{2}{c}{p-boot} &\multicolumn{2}{c}{p-boot} & \multicolumn{2}{c}{np-boot}& \multicolumn{2}{c}{np-boot}\\
     & $\hat{\mu}_i$        &MSE$\times 10^5$&  $\tilde{\mu}_i$        &MSE$\times 10^5$& $\overline{\mu}_{1,i}$        &MSE$\times 10^5$& $\overline{\overline{\mu}}_{1,i}$        &MSE$\times 10^5$& $\overline{\mu}_{2,i}$        &MSE$\times 10^5$& $\overline{\overline{\mu}}_{2,i}$        &MSE$\times 10^5$\\
    \hline
 $21$&0.97785&2.4852&0.97815&2.4196&0.97816&2.4197&0.97811&2.4307&0.97846&2.5379&0.97810&2.6007 \\
 $22$&0.97141&1.3669&0.97149&1.3634&0.97146&1.3661&0.97144&1.3680&0.97157&1.3982&0.97148&1.4085 \\
 $23$&0.98232&0.4331&0.98234&0.4330&0.98233&0.4326&0.98234&0.4323&0.98241&0.4570&0.98244&0.4585 \\
 $24$&0.97211&1.2809&0.97219&1.2778&0.97215&1.2802&0.97214&1.2820&0.97226&1.3105&0.97218&1.3200 \\
 $25$&0.91094&2.2236&0.91144&2.2168&0.91131&2.2203&0.91121&2.2236&0.91175&2.2951&0.91107&2.3228 \\
 $26$&0.99623&0.0233&0.99623&0.0233&0.99624&0.0233&0.99624&0.0232&0.99628&0.0256&0.99628&0.0252 \\
 $27$&0.98186&0.2087&0.98184&0.2100&0.98183&0.2100&0.98183&0.2100&0.98187&0.2143&0.98192&0.2166 \\
 $28$&0.98579&0.1580&0.98578&0.1584&0.98577&0.1586&0.98577&0.1587&0.98582&0.1634&0.98584&0.1646 \\
 $29$&0.99109&0.1230&0.99110&0.1226&0.99110&0.1227&0.99110&0.1228&0.99117&0.1312&0.99115&0.1313 \\
 $30$&0.98713&0.1437&0.98713&0.1442&0.98712&0.1442&0.98713&0.1439&0.98717&0.1511&0.98722&0.1522 \\
 $31$&0.95757&6.8724&0.95797&6.7758&0.95793&6.7897&0.95786&6.8087&0.95827&7.0365&0.95781&7.1589 \\
 $32$&0.98757&0.3547&0.98763&0.3517&0.98763&0.3519&0.98762&0.3526&0.98773&0.3718&0.98766&0.3745 \\
 $33$&0.95759&2.9853&0.95771&2.9838&0.95765&2.9856&0.95763&2.9892&0.95780&3.0718&0.95767&3.0980 \\
 $34$&0.97058&1.3386&0.97065&1.3367&0.97062&1.3392&0.97060&1.3409&0.97072&1.3674&0.97064&1.3766 \\
 $35$&0.97648&1.5266&0.97663&1.5100&0.97661&1.5124&0.97659&1.5161&0.97678&1.5722&0.97661&1.5928 \\
 $36$&0.96553&1.6061&0.96559&1.6084&0.96555&1.6088&0.96553&1.6103&0.96565&1.6551&0.96560&1.6669 \\
 $37$&0.97824&0.9721&0.97833&0.9660&0.97831&0.9678&0.97829&0.9695&0.97843&1.0024&0.97833&1.0117 \\
 $38$&0.88094&6.2588&0.88213&6.1947&0.88206&6.1936&0.88186&6.2063&0.88316&6.5824&0.88165&6.7035 \\
 $39$&0.92631&1.4425&0.92671&1.4377&0.92660&1.4408&0.92652&1.4429&0.92693&1.4795&0.92641&1.4951 \\
 $40$&0.97234&0.8019&0.97236&0.8047&0.97233&0.8049&0.97233&0.8054&0.97240&0.8246&0.97240&0.8298 \\
\hline
\end{tabular}
\end{center}
\label{tabelasimulmu22}
\end{sidewaystable}

\begin{sidewaystable}
\caption{Estimated values of $\phi$ for $n=40$ for observations $1$ until $20$.}
\begin{center}
\begin{tabular}{ccccccccccccccccc}
\hline
$i$ &True & \multicolumn{2}{c}{MLE}& \multicolumn{2}{c}{Cox-Snell}  & \multicolumn{2}{c}{p-boot} & \multicolumn{2}{c}{np-boot}\\
&$\phi_i$ &  $\hat{\phi}_i$       &MSE$\times 10^{-5}$&  $\tilde{\phi}$        &MSE$\times 10^{-5}$& $\overline{\overline{\phi}}_{1,i}$        &MSE$\times 10^{-5}$& $\overline{\overline{\phi}}_{2,i}$       &MSE$\times 10^{-5}$\\
    \hline
 $1$&1770.8&3029.1&68.054&1861.3&15.910&1663.4&10.169&2125.6&29.674 \\
 $2$&959.67&1421.0&9.6333&1001.7&3.4732&951.83&3.0914&1089.0&5.3961 \\
 $3$&2941.7&6339.4&692.50&2939.7&81.138&2361.6&34.781&3536.2&213.27 \\
 $4$&60.601&73.479&0.0123&61.313&0.0068&61.061&0.0069&66.301&0.0098 \\
 $5$&257.67&347.87&0.3260&265.33&0.1327&256.31&0.1191&294.61&0.2089 \\
 $6$&239.17&308.30&0.1628&245.17&0.0696&238.31&0.0642&269.20&0.1141 \\
 $7$&215.80&266.89&0.0776&219.99&0.0345&215.29&0.0336&237.73&0.0616 \\
 $8$&174.37&214.83&0.0531&177.46&0.0246&174.06&0.0240&192.51&0.0424 \\
 $9$&1201.1&1996.5&27.074&1258.5&6.7785&1140.1&4.6161&1437.9&12.362 \\
$10$&42.483&50.631&0.0065&42.945&0.0037&42.900&0.0037&43.995&0.0044 \\
$11$&445.50&598.31&0.7136&460.68&0.2667&441.88&0.2294&507.42&0.4530 \\
$12$&56.060&67.187&0.0094&56.627&0.0054&56.471&0.0054&60.968&0.0077 \\
$13$&45.972&55.403&0.0086&46.559&0.0049&46.466&0.0048&47.623&0.0056 \\
$14$&5615.7&11685.&2277.4&5774.4&324.04&4721.4&133.08&6740.9&644.14 \\
$15$&355.04&458.57&0.3073&364.99&0.1224&353.29&0.1115&397.95&0.2142 \\
$16$&12.357&14.273&0.0006&12.431&0.0004&12.546&0.0003&12.715&0.0004 \\
$17$&32.844&42.080&0.0102&33.430&0.0051&33.534&0.0051&36.444&0.0069 \\
$18$&215.87&267.12&0.0936&220.10&0.0442&215.76&0.0435&235.81&0.0741 \\
$19$&10.734&13.121&0.0011&10.906&0.0006&11.058&0.0006&11.452&0.0007 \\
$20$&13.496&17.255&0.0027&13.777&0.0013&13.960&0.0013&14.747&0.0016 \\
\hline
\end{tabular}
\end{center}
\label{tabelasimulphi21}
\end{sidewaystable}

\begin{sidewaystable}
\caption{Estimated values of $\phi$ for $n=40$ for observations $21$ until $40$.}
\begin{center}
\begin{tabular}{ccccccccccccccccc}
\hline
$i$ &True & \multicolumn{2}{c}{MLE}& \multicolumn{2}{c}{Cox-Snell}  & \multicolumn{2}{c}{p-boot} & \multicolumn{2}{c}{np-boot}\\
&$\phi_i$ &  $\hat{\phi}_i$       &MSE$\times 10^{-5}$&  $\tilde{\phi}$        &MSE$\times 10^{-5}$& $\overline{\overline{\phi}}_{1,i}$        &MSE$\times 10^{-5}$& $\overline{\overline{\phi}}_{2,i}$       &MSE$\times 10^{-5}$\\
    \hline
 $21$&98.278&139.68&0.1618&101.37&0.0661&100.08&0.0637&102.30&0.0745 \\
 $22$&81.514&96.913&0.0154&82.397&0.0088&81.888&0.0088&86.262&0.0123 \\
 $23$&438.48&646.61&1.8109&455.33&0.6103&430.67&0.5010&513.31&0.9936 \\
 $24$&85.347&101.64&0.0168&86.300&0.0095&85.725&0.0095&90.428&0.0135 \\
 $25$&12.081&13.979&0.0006&12.158&0.0004&12.277&0.0003&12.462&0.0004 \\
 $26$&6878.7&14673.&4140.1&7032.1&562.61&5693.9&216.64&8204.3&1048.4 \\
 $27$&282.12&361.31&0.1870&289.31&0.0770&280.89&0.0708&316.26&0.1327 \\
 $28$&352.09&452.61&0.3283&361.66&0.1399&351.18&0.1318&390.82&0.2357 \\
 $29$&751.25&1065.4&4.1163&781.77&1.5804&747.43&1.4258&848.19&2.4693 \\
 $30$&769.52&1177.0&6.4918&804.56&1.9558&748.37&1.4798&908.18&3.3472 \\
 $31$&35.909&45.474&0.0098&36.618&0.0050&36.509&0.0048&36.669&0.0051 \\
 $32$&339.05&458.89&0.8045&349.84&0.3545&340.78&0.3397&371.61&0.5103 \\
 $33$&48.458&56.311&0.0048&48.732&0.0029&48.711&0.0029&51.637&0.0041 \\
 $34$&79.804&94.260&0.0129&80.570&0.0074&80.097&0.0074&84.581&0.0106 \\
 $35$&100.14&126.94&0.0572&102.25&0.0293&101.28&0.0288&105.67&0.0367 \\
 $36$&74.761&88.166&0.0098&75.391&0.0056&74.957&0.0057&80.671&0.0087 \\
 $37$&123.37&153.22&0.0590&125.66&0.0310&124.25&0.0306&131.42&0.0421 \\
 $38$&8.7913&11.450&0.0016&9.0134&0.0007&9.1672&0.0007&9.5762&0.0008 \\
 $39$&15.961&18.414&0.0009&16.046&0.0005&16.152&0.0005&16.341&0.0006 \\
 $40$&115.31&138.36&0.0209&116.72&0.0108&115.33&0.0109&125.69&0.0181 \\
\hline
\end{tabular}
\end{center}
\label{tabelasimulphi22}
\end{sidewaystable}

\begin{table}[t!]
\caption{Simulation results for $n=60$.}
\begin{center}
\begin{tabular}{lccccc}
\hline
Parameter & MLE& Cox-Snell & Firth & p-boot & np-boot\\
\hline
$\beta_0$   & 1.5046   &1.5025  &1.5063   &1.5016   &1.5005   \\
Bias        & 0.0046   &0.0025  &0.0063   &0.0016   &0.0005          \\ 
Variance    & 0.0156   &0.0156  &0.0159   &0.0156   &0.0160 \\
MSE         & 0.0156   &0.0156  &0.0159   &0.0156   &0.0160 \\\\
$\beta_1$   & 0.4995   &0.4996  &0.4996   &0.4997   &0.5000         \\
Bias        &-0.0005   &-0.0004 &-0.0004  &-0.0003  &0.0000                \\
Variance    & 0.0030   &0.0030  &0.0031   &0.0030   &0.0033  \\
MSE         & 0.0030   &0.0030  &0.0031   &0.0030   &0.0033\\\\
$\beta_2$   & 1.9984   &1.9992  &1.9976   &1.9997   &2.0014   \\
Bias        &-0.0016   &0.0008  &-0.0024  &-0.0003  &0.0014               \\
Variance    & 0.0037   &0.0037  &0.0038   &0.0037   &0.0039\\
MSE         & 0.0037   &0.0037  &0.0038   &0.0037   &0.0039\\\\
$\theta_0$  & 1.7004   &1.6948  &1.6939   &1.7089   &1.7066   \\
Bias        & 0.0004   &-0.0052 &-0.0061  &0.0089   &0.0066               \\
Variance    & 0.0999   &0.0962  &0.1059   &0.0956   &0.0993 \\
MSE         & 0.0999   &0.0962  &0.1059   &0.0957   &0.0994 \\\\
$\theta_1$  & 0.6966   &0.6974  &0.6955   &0.6988   &0.7015  \\
Bias        &-0.0034   &-0.0026 &-0.0045  &-0.0012  &0.0015                \\
Variance    & 0.0589   &0.0551  &0.0650   &0.0543   &0.0581 \\
MSE         & 0.0589   &0.0551  &0.0650   &0.0543   &0.0581 \\\\
$\theta_2$  & 3.0630   &3.0107  &3.1184   &2.9955   &3.0073   \\
Bias        & 0.0630   &0.0107  &0.1184   &-0.0045  &0.0073               \\
Variance    & 0.0265   &0.0260  &0.0274   &0.0257   &0.0271 \\
MSE         & 0.0305   &0.0262  &0.0414   &0.0257   &0.0271 \\
\hline
\end{tabular}
\end{center}
\label{resulsimul3}
\end{table}

\begin{sidewaystable}
\caption{Estimated values of $\mu$ for $n=60$ for observations $1$ until $30$.}
\begin{center}
\begin{tabular}{cccccccccccccccc}
\hline
$i$ & \multicolumn{2}{c}{MLE}& \multicolumn{2}{c}{Cox-Snell}  & \multicolumn{2}{c}{p-boot} &\multicolumn{2}{c}{p-boot} & \multicolumn{2}{c}{np-boot}& \multicolumn{2}{c}{np-boot}\\
     & $\hat{\mu}_i$        &MSE$\times 10^5$&  $\tilde{\mu}_i$        &MSE$\times 10^5$& $\overline{\mu}_{1,i}$        &MSE$\times 10^5$& $\overline{\overline{\mu}}_{1,i}$        &MSE$\times 10^5$& $\overline{\mu}_{2,i}$        &MSE$\times 10^5$& $\overline{\overline{\mu}}_{2,i}$        &MSE$\times 10^5$\\
    \hline
 $1$&0.94851&3.0577&0.94865&3.0528&0.94863&3.0590&0.94861&3.0605&0.94867&3.1232&0.94857&3.1344 \\
 $2$&0.98379&0.3952&0.98386&0.3922&0.98386&0.3927&0.98385&0.3931&0.98392&0.4095&0.98385&0.4123 \\
 $3$&0.98630&0.0577&0.98629&0.0578&0.98629&0.0579&0.98629&0.0579&0.98630&0.0594&0.98631&0.0595 \\
 $4$&0.94256&4.0143&0.94272&4.0065&0.94270&4.0161&0.94268&4.0182&0.94279&4.2206&0.94264&4.2429 \\
 $5$&0.90825&1.1968&0.90856&1.1956&0.90853&1.1985&0.90849&1.1990&0.90865&1.2481&0.90837&1.2543 \\
 $6$&0.99074&0.0254&0.99073&0.0254&0.99073&0.0255&0.99073&0.0255&0.99074&0.0265&0.99075&0.0266 \\
 $7$&0.98393&0.1028&0.98393&0.1029&0.98393&0.1031&0.98393&0.1031&0.98394&0.1044&0.98394&0.1045 \\
 $8$&0.97812&0.2439&0.97813&0.2442&0.97812&0.2446&0.97812&0.2447&0.97814&0.2504&0.97813&0.2507 \\
 $9$&0.97980&0.1898&0.97980&0.1900&0.97979&0.1904&0.97979&0.1905&0.97981&0.1952&0.97981&0.1954 \\
$10$&0.99198&0.0193&0.99198&0.0193&0.99197&0.0194&0.99197&0.0194&0.99199&0.0200&0.99199&0.0201 \\
$11$&0.92807&6.7434&0.92829&6.7353&0.92826&6.7497&0.92823&6.7528&0.92832&6.9123&0.92816&6.9388 \\
$12$&0.95979&1.5525&0.95987&1.5509&0.95986&1.5545&0.95985&1.5552&0.95992&1.6338&0.95984&1.6405 \\
$13$&0.97648&0.4977&0.97653&0.4968&0.97652&0.4977&0.97651&0.4979&0.97655&0.5077&0.97652&0.5091 \\
$14$&0.97879&0.4744&0.97884&0.4728&0.97884&0.4736&0.97883&0.4739&0.97887&0.4862&0.97883&0.4881 \\
$15$&0.96264&1.3531&0.96272&1.3514&0.96271&1.3540&0.96269&1.3546&0.96273&1.3791&0.96268&1.3831 \\
$16$&0.96332&1.1294&0.96338&1.1291&0.96336&1.1316&0.96335&1.1320&0.96340&1.1746&0.96335&1.1785 \\
$17$&0.92130&1.0687&0.92164&1.0642&0.92163&1.0670&0.92158&1.0677&0.92189&1.1409&0.92150&1.1489 \\
$18$&0.96480&1.3207&0.96489&1.3178&0.96488&1.3203&0.96486&1.3210&0.96491&1.3447&0.96485&1.3491 \\
$19$&0.86528&3.2908&0.86585&3.2893&0.86580&3.2987&0.86573&3.2998&0.86613&3.5018&0.86551&3.5203 \\
$20$&0.94079&5.9637&0.94105&5.9313&0.94104&5.9464&0.94101&5.9508&0.94128&6.4000&0.94097&6.4448 \\
$21$&0.99231&0.0820&0.99234&0.0814&0.99234&0.0815&0.99233&0.0816&0.99238&0.0868&0.99234&0.0873 \\
$22$&0.99659&0.0056&0.99659&0.0056&0.99659&0.0056&0.99659&0.0056&0.99660&0.0060&0.99660&0.0059 \\
$23$&0.96626&0.9311&0.96630&0.9309&0.96629&0.9330&0.96629&0.9334&0.96633&0.9773&0.96629&0.9805 \\
$24$&0.95962&1.8661&0.95974&1.8617&0.95972&1.8652&0.95971&1.8662&0.95976&1.9004&0.95969&1.9071 \\
$25$&0.99291&0.0245&0.99290&0.0245&0.99291&0.0245&0.99291&0.0245&0.99293&0.0258&0.99293&0.0258 \\
$26$&0.98481&0.2667&0.98485&0.2654&0.98485&0.2658&0.98484&0.2660&0.98489&0.2761&0.98485&0.2774 \\
$27$&0.97509&0.3725&0.97511&0.3728&0.97510&0.3735&0.97510&0.3737&0.97512&0.3811&0.97510&0.3818 \\
$28$&0.99005&0.0375&0.99004&0.0375&0.99004&0.0376&0.99004&0.0376&0.99006&0.0387&0.99006&0.0387 \\
$29$&0.99756&0.0053&0.99756&0.0053&0.99756&0.0053&0.99756&0.0053&0.99758&0.0058&0.99757&0.0058 \\
$30$&0.99502&0.0122&0.99502&0.0122&0.99502&0.0122&0.99502&0.0122&0.99503&0.0128&0.99503&0.0128 \\
\hline
\end{tabular}
\end{center}
\label{tabelasimulmu31}
\end{sidewaystable}
\begin{sidewaystable}
\caption{Estimated values of $\mu$ for $n=60$ for observations $31$ until $60$.}
\begin{center}
\begin{tabular}{cccccccccccccccc}
\hline
$i$ & \multicolumn{2}{c}{MLE}& \multicolumn{2}{c}{Cox-Snell}  & \multicolumn{2}{c}{p-boot} &\multicolumn{2}{c}{p-boot} & \multicolumn{2}{c}{np-boot}& \multicolumn{2}{c}{np-boot}\\
     & $\hat{\mu}_i$        &MSE$\times 10^5$&  $\tilde{\mu}_i$        &MSE$\times 10^5$& $\overline{\mu}_{1,i}$        &MSE$\times 10^5$& $\overline{\overline{\mu}}_{1,i}$        &MSE$\times 10^5$& $\overline{\mu}_{2,i}$        &MSE$\times 10^5$& $\overline{\overline{\mu}}_{2,i}$        &MSE$\times 10^5$\\
    \hline
 $31$&0.99415&0.0184&0.99415&0.0184&0.99415&0.0185&0.99415&0.0185&0.99417&0.0197&0.99416&0.0197 \\
 $32$&0.99216&0.0559&0.99217&0.0557&0.99217&0.0558&0.99217&0.0558&0.99220&0.0591&0.99218&0.0593 \\
 $33$&0.97738&0.2732&0.97739&0.2736&0.97738&0.2741&0.97738&0.2742&0.97740&0.2838&0.97739&0.2842 \\
 $34$&0.98368&0.1268&0.98369&0.1270&0.98368&0.1272&0.98368&0.1272&0.98369&0.1289&0.98369&0.1290 \\
 $35$&0.97015&1.2755&0.97026&1.2683&0.97026&1.2705&0.97024&1.2715&0.97032&1.3055&0.97023&1.3126 \\
 $36$&0.94593&3.2106&0.94606&3.2069&0.94604&3.2139&0.94602&3.2155&0.94609&3.3192&0.94598&3.3329 \\
 $37$&0.95951&2.3624&0.95966&2.3515&0.95965&2.3557&0.95963&2.3574&0.95972&2.4107&0.95960&2.4225 \\
 $38$&0.93229&8.6609&0.93261&8.6077&0.93261&8.6301&0.93257&8.6368&0.93291&9.3024&0.93251&9.3723 \\
 $39$&0.93427&5.3243&0.93446&5.3179&0.93444&5.3293&0.93441&5.3319&0.93449&5.4728&0.93435&5.4948 \\
 $40$&0.96087&1.3510&0.96094&1.3505&0.96092&1.3532&0.96092&1.3538&0.96095&1.3912&0.96090&1.3956 \\
 $41$&0.98101&0.1971&0.98102&0.1972&0.98102&0.1975&0.98102&0.1976&0.98103&0.2003&0.98103&0.2005 \\
 $42$&0.98213&0.1386&0.98213&0.1388&0.98213&0.1390&0.98213&0.1391&0.98214&0.1410&0.98214&0.1411 \\
 $43$&0.98439&0.1327&0.98440&0.1327&0.98440&0.1329&0.98439&0.1330&0.98441&0.1353&0.98441&0.1355 \\
 $44$&0.98581&0.0634&0.98580&0.0635&0.98579&0.0636&0.98579&0.0636&0.98581&0.0652&0.98581&0.0653 \\
 $45$&0.96663&1.0672&0.96670&1.0656&0.96669&1.0676&0.96668&1.0681&0.96671&1.0864&0.96667&1.0895 \\
 $46$&0.97528&0.4526&0.97531&0.4524&0.97530&0.4532&0.97529&0.4534&0.97532&0.4604&0.97530&0.4613 \\
 $47$&0.99004&0.0224&0.99003&0.0224&0.99003&0.0224&0.99003&0.0224&0.99004&0.0228&0.99004&0.0229 \\
 $48$&0.98474&0.0802&0.98474&0.0803&0.98473&0.0804&0.98473&0.0804&0.98474&0.0817&0.98475&0.0817 \\
 $49$&0.99354&0.0221&0.99354&0.0221&0.99354&0.0221&0.99354&0.0221&0.99356&0.0235&0.99355&0.0235 \\
 $50$&0.98988&0.1142&0.98989&0.1138&0.98990&0.1139&0.98990&0.1140&0.98995&0.1220&0.98992&0.1217 \\
 $51$&0.98646&0.0657&0.98645&0.0658&0.98645&0.0659&0.98645&0.0659&0.98646&0.0667&0.98647&0.0667 \\
 $52$&0.92288&7.7763&0.92312&7.7688&0.92309&7.7860&0.92306&7.7895&0.92316&8.0112&0.92297&8.0444 \\
 $53$&0.94860&3.0225&0.94873&3.0171&0.94872&3.0242&0.94870&3.0258&0.94880&3.1809&0.94867&3.1969 \\
 $54$&0.98109&0.2037&0.98110&0.2038&0.98109&0.2042&0.98109&0.2043&0.98111&0.2071&0.98110&0.2074 \\
 $55$&0.95698&1.8418&0.95708&1.8398&0.95706&1.8435&0.95705&1.8444&0.95709&1.8839&0.95702&1.8899 \\
 $56$&0.94157&4.1167&0.94173&4.1093&0.94171&4.1190&0.94169&4.1212&0.94180&4.3166&0.94164&4.3388 \\
 $57$&0.95495&2.0132&0.95505&2.0114&0.95503&2.0156&0.95502&2.0165&0.95506&2.0675&0.95499&2.0745 \\
 $58$&0.99070&0.0199&0.99069&0.0199&0.99069&0.0200&0.99069&0.0200&0.99070&0.0205&0.99071&0.0206 \\
 $59$&0.98021&0.2164&0.98021&0.2166&0.98021&0.2171&0.98021&0.2172&0.98024&0.2291&0.98022&0.2294 \\
 $60$&0.96732&0.8075&0.96736&0.8076&0.96735&0.8092&0.96734&0.8096&0.96737&0.8327&0.96734&0.8349 \\
\hline
\end{tabular}
\end{center}
\label{tabelasimulmu32}
\end{sidewaystable}

\begin{sidewaystable}
\caption{Estimated values of $\phi$ for $n=60$ for observations $1$ until $30$.}
\begin{center}
\begin{tabular}{ccccccccccccccccc}
\hline
$i$ &True & \multicolumn{2}{c}{MLE}& \multicolumn{2}{c}{Cox-Snell}  & \multicolumn{2}{c}{p-boot} & \multicolumn{2}{c}{np-boot}\\
&$\phi_i$ &  $\hat{\phi}_i$       &MSE$\times 10^{-5}$&  $\tilde{\phi}$        &MSE$\times 10^{-5}$& $\overline{\overline{\phi}}_{1,i}$        &MSE$\times 10^{-5}$& $\overline{\overline{\phi}}_{2,i}$       &MSE$\times 10^{-5}$\\
    \hline
 $1$&28.336&30.011&0.0009&28.460&0.0007&28.793&0.0007&28.813&0.0008 \\
 $2$&183.72&214.70&0.1450&186.54&0.0940&187.63&0.0943&190.42&0.1049 \\
 $3$&623.86&783.08&0.8483&637.06&0.3770 &615.45&0.3434&637.20&0.4139 \\
 $4$&29.638&32.969&0.0015&30.024&0.0011&30.101&0.0011&30.249&0.0012 \\
 $5$&11.385&12.224&0.0002&11.496&0.0001&11.637&0.0001&11.607&0.0001 \\
 $6$&1936.2&2820.9& 31.510&1994.3&10.719&1858.5&8.7753&1975.3&12.579 \\
 $7$&334.56&388.42&0.1007&338.44&0.0540&333.00&0.0525&340.76&0.0577 \\
 $8$&190.37&217.33&0.0290&192.45&0.0169&190.15&0.0166&193.66&0.0179 \\
 $9$&234.57&271.95&0.0512&237.57&0.0280&233.85&0.0272&238.80&0.0299 \\
$10$&1495.4&1929.1& 6.7590&1524.6&2.8932&1466.1&2.6041&1530.2&3.1911 \\
$11$&16.194&17.057&0.0003&16.274&0.0002&16.507&0.0002&16.471&0.0003 \\
$12$&62.674&71.139&0.0060&63.538&0.0040&63.215&0.0040&63.914&0.0043 \\
$13$&112.79&123.84&0.0171&113.52&0.0126&113.95&0.0127&115.12&0.0136 \\
$14$&127.31&141.69&0.0321&128.41&0.0231&129.05&0.0232&130.48&0.0251 \\
$15$&50.908&54.476&0.0025&51.129&0.0020&51.519&0.0020&51.756&0.0021 \\
$16$&67.847&75.284&0.0046&68.498&0.0032&68.332&0.0032&69.005&0.0034 \\
$17$&18.741&22.188&0.0016&19.207&0.0010&19.210&0.0010&19.326&0.0011 \\
$18$&53.274&57.218&0.0037&53.553&0.0029&54.055&0.0030&54.296&0.0031 \\
$19$&6.1804&7.0164&0.0001&6.3181&0.0001&6.3891&0.0001&6.3720&0.0001 \\
$20$&37.974&47.434&0.0085&39.074&0.0046&38.636&0.0045&39.198&0.0051 \\
$21$&840.75&1063.0&4.2129&854.48&2.2201&844.57&2.1370 &872.19&2.5198 \\
$22$&17470 &33191 & 14821&17324 &2650.2&14743 &1637.7&16942 &3861.7 \\
$23$&88.824&101.36&0.0105&90.021&0.0067&89.288&0.0066&90.512&0.0072 \\
$24$&41.464&44.323&0.0023&41.681&0.0019&42.147&0.0019&42.262&0.0020 \\
$25$&4942.5&8778.9&796.04&5024.8&175.69&4406.4&120.53&4904.7&237.83 \\
$26$&228.26&263.24&0.1540&231.01&0.1014&231.32&0.1009&235.17&0.1122 \\
$27$&133.92&149.16&0.0122&134.97&0.0080&134.19&0.0080&136.09&0.0085 \\
$28$&834.22&1018.7& 1.2950&846.76&0.6351&825.35&0.5958&852.99&0.6903 \\
$29$&23640 &43563 & 22547&23263 &4383.7&20224 &2945.3&23053 &6021.5 \\
$30$&10969 &21679 &7431.4&10862 &1203.8&9077.1&705.40 &10524 &1851.2 \\
\hline
\end{tabular}
\end{center}
\label{tabelasimulphi31}
\end{sidewaystable}

\begin{sidewaystable}
\caption{Estimated values of $\phi$ for $n=60$ for observations $31$ until $60$.}
\begin{center}
\begin{tabular}{ccccccccccccccccc}
\hline
$i$ &True & \multicolumn{2}{c}{MLE}& \multicolumn{2}{c}{Cox-Snell}  & \multicolumn{2}{c}{p-boot} & \multicolumn{2}{c}{np-boot}\\
   &$\phi_i$ &  $\hat{\phi}_i$       &MSE$\times 10^{-5}$&  $\tilde{\phi}$        &MSE$\times 10^{-5}$& $\overline{\overline{\phi}}_{1,i}$        &MSE$\times 10^{-5}$& $\overline{\overline{\phi}}_{2,i}$       &MSE$\times 10^{-5}$\\
    \hline
 $31$&2465.2&3299.1&29.664&2510.6&12.348&2405.9&11.019&2528.1&13.718  \\
 $32$&975.65&1214.3&3.6737&990.08&1.9334&972.07&1.8371&1005.3&2.1481  \\
 $33$&197.61&230.03&0.0428&200.38&0.0239&197.23&0.0231&201.27&0.0255 \\
 $34$&290.07&331.45&0.0744&292.80&0.0439&289.50&0.0432&295.47&0.0471 \\
 $35$&63.746&70.602&0.0114&64.452&0.0084&65.238&0.0085&65.605&0.0091 \\
 $36$&28.936&31.051&0.0009&29.124&0.0007&29.343&0.0007&29.410&0.0007 \\
 $37$&38.716&42.101&0.0035&39.078&0.0027&39.639&0.0028&39.728&0.0029 \\
 $38$&29.493&37.284&0.0061&30.411&0.0033&30.111&0.0032&30.521&0.0037 \\
 $39$&19.193&20.294&0.0004&19.291&0.0003&19.530&0.0003&19.509&0.0003 \\
 $40$&53.662&58.158&0.0024&53.994&0.0018&54.150&0.0018&54.503&0.0019 \\
 $41$&208.25&233.97&0.0344&209.88&0.0219&208.39&0.0218&211.95&0.0234 \\
 $42$&270.23&310.60&0.0589&273.13&0.0328&269.36&0.0321&275.05&0.0349 \\
 $43$&293.62&334.80&0.0908&296.24&0.0560&293.51&0.0551&299.42&0.0604 \\
 $44$&557.66&690.46&0.5811&568.65&0.2645&551.07&0.2434&569.32&0.2888 \\
 $45$&61.054&65.632&0.0040&61.340&0.0031&61.781&0.0032&62.133&0.0033 \\
 $46$&113.76&124.50&0.0113&114.39&0.0082&114.48&0.0083&115.74&0.0088 \\
 $47$&1185.9&1546.1&4.3867&1213.1&1.7976&1161.3&1.5920&1212.3&2.0038 \\
 $48$&416.39&495.70&0.2017&422.59&0.0993&413.21&0.0944&424.45&0.1068 \\
 $49$&1942.0  &2538.4&15.298&1976.5&6.7135&1904.3&6.0813&1992.1&7.4153 \\
 $50$&3384.0  &6771.7&742.04&3342.6&120.27&2812.3&74.335&3228.8&185.45 \\
 $51$&470.75&556.27&0.2506&476.88&0.1281&467.43&0.1228&480.04&0.1378 \\
 $52$&14.686&15.511&0.0002&14.770&0.0002&14.970&0.0002&14.934&0.0002 \\
 $53$&37.545&42.075&0.0024&38.050&0.0017&38.052&0.0017&38.315&0.0018 \\
 $54$&204.71&229.58&0.0349&206.26&0.0226&204.97&0.0225&208.40&0.0242 \\
 $55$&40.884&43.632&0.0015&41.066&0.0012&41.387&0.0012&41.531&0.0013 \\
 $56$&27.923&30.801&0.0013&28.251&0.0009&28.362&0.0009&28.475&0.0010 \\
 $57$&39.067&41.825&0.0014&39.266&0.0011&39.533&0.0011&39.678&0.0012 \\
 $58$&1604.0 &2205.3&13.224&1647.4&4.9603&1557.0&4.2277&1639.4&5.6607 \\
 $59$&327.94&410.19&0.2761&335.44&0.1322&325.32&0.1216&335.25&0.1439 \\
 $60$&80.047&88.150&0.0050&80.661&0.0036&80.508&0.0036&81.329&0.0038 \\

\hline
\end{tabular}
\end{center}
\label{tabelasimulphi32}
\end{sidewaystable}

In each of the $5,000$ replications, we fitted the model and computed the MLEs $\hat{\beta}$,
$\hat{\theta}$, its corrected versions from the corrective method (Cox and Snell, 1968),
preventive method (Firth, 1993) and the bootstrap method both of its parametric and nonparametric versions (Efron, 1979).
The number of bootstrap replications was set to 600 for both bootstrap methods.

In order to analyze the results we computed, for each sample size and for each estimator, the mean of estimates, bias, variance and mean square error (MSE). Tables \ref{resulsimul}-\ref{tabelasimulphi32} present simulation results for sample sizes $n=20,40$ and $60$, respectively. Moreover, we also considered the estimated values of $\mu_i$ and $\phi_i$, $i = 1,\ldots, n$, $n=20,40$ and $60$, along with its corrected estimators presented in Section \ref{biasmuphi}. We also want to emphasize that, for the bootstrap schemes, two methods to removing the first-order bias can be considered: the one induced by the Taylor's expansion and the one given by the bootstrap itself, denoted by $\overline{\alpha}_i$ and $\overline{\overline{\alpha}}_i$, respectively, where $\alpha_i$ is a surrogate for $\mu_i$ and $\phi_i$. 

Table \ref{resulsimul} presents simulation results for sample size $n=20$ with respect to the parameters $\beta$ and $\theta$. We begin by looking the estimated biases, in absolute value, of the estimators. Initially we note that for all parameters the biases of the corrective estimators were smaller than of the original MLEs. However, for all parameters the biases of preventive estimators were bigger than of the original MLEs, moreover, not only the biases were bigger but also the MSE, which shows that the preventive method does not work well for this model, the same phenomenon occurred in Ospina et al. (2006), which collaborates to the idea that this method has some problems in beta regression models. We now observe that the MSE of the corrective estimators were smaller than those of the MLEs for all parameters, showing that the correction is effective. Moving to the bootstrap corrected-estimators we note that the parametric bootstrap had the smallest MSE for all parameters, even though the biases were not the smallest. However, the MSE were very close to the MSE of the corrective method, and the computation of the parametric bootstrap biases is computer intensive, whereas the corrective method is not. Lastly, we observe that for all parameters $\theta$ the MSE of the nonparametric bootstrap corrected estimators were smaller than those of the MLEs. Moreover, for the parameters $\beta$, the MSE of the nonparametric bootstrap corrected estimators were very close to those of the MLEs, showing that this method is satisfactory, and is very easy to implement by practioners since no parametric assumptions are made. Therefore, for the small sample size $n=20$, we were able to conclude that corrective method by Cox and Snell (1969) was successfully applied, as well as the bootstrap correction.

Table \ref{tabelasimulmu1} presents the simulation results for sample size $n=20$ with respect to the parameter $\mu$. We did not consider the preventive estimators since its estimation was poor. The MSE of both nonparametric bootstrap correction schemes are the biggest ones for all parameters. The best estimators, with respect to MSE, are the parametric bootstrap schemes. Finally, the MSE of the corrective method are always least or equal those of the MLE. Note also that, for the for the means, all the values are not distant from each other. These results show that, even for small sample, we are able to obtain a satisfactory bias correction for both the corrective method and the parametric bootstrap method.

Table \ref{tabelasimulphi1} presents the simulation results for sample size $n=20$ with respect to the parameter $\phi$. The bootstrap correction scheme based on Taylor's expasion is not presented in this Table, since its performance was worse than those of the regular bootstrap correction, for both parametric and nonparametric schemes. The first remarkable fact is that if the true value of $\phi$ is large, the bias of the corrective method is small when compared with the others estimators, and the bias of the parametric bootstrap scheme is considerably larger than the ones of the corrective method. If the true value of $\phi$ is small, the parametric bootstrap outperforms the corrective method with respect to bias. The maximum likelihood estimator had the worst perfomance with respect to bias and MSE. The parametric bootstrap had, in general, the best performance with respect to MSE. Therefore, we conclude that the corrective method was satisfactory, but overall, the parametric bootstrap had the best performance. It is also noteworthy that, for the precision parameter $\phi$, the correction schemes worked very well, and, therefore, their use produces an improved estimation.

In Table \ref{resulsimul2} the results for sample size $n=40$ are presented with respect to the parameters $\beta$ and $\theta$. As observed in Table \ref{resulsimul}, the corrective method and the parametric bootstrap scheme had the best performance with respect to both bias and MSE. The nonparametric bootstrap scheme had performance worse than the MLE for the parameters $\beta$, and better than the MLE for the parameters $\theta$, with respect to both bias and MSE. For this sample size, we observe again that the preventive estimator had a poor performance. The results show that both the corrective method and the parametric bootstrap produce better estimators than the MLE, with respect to bias and MSE, and again, that the estimates produced by the preventive method are not satisfactory. Finally, the behavior of the nonparametric bootstrap estimator is the same as the one for sample size $n=20$.

Tables \ref{tabelasimulmu21} and \ref{tabelasimulmu22} contains the results for sample size $n=40$ with respect to the parameter $\mu$. Initially we note that, based on MSE, the corrective estimators had the best performance. We also observe that the maximum likelihood and parametric bootstrap estimators had similar performance. Finally, the nonparametric estimators had the worst performance with regard to MSE.

Tables \ref{tabelasimulphi21} and \ref{tabelasimulphi22} sumarize the results for sample size $n=40$ with respect to the parameter $\phi$. For this sample size the MLE had the biggest bias and MSE, and hence was the poorest estimator considered in this table. On the opposite direction, the parametric bootstrap estimator had the smallest MSE, followed by the corrective method. However, if the true value is large, the corrective method has bias smaller than the parametric bootstrap. It is also noteworthy that for large values of $\phi$ the parametric bootstrap method tends to underestimate the parameter. We note, again, that both the corrective and parametric bootstrap estimators provide better results than the MLE, with respect to both bias and MSE.

Table \ref{resulsimul3} summarizes the results for sample size $n=60$ with respect to the parameters $\beta$ and $\theta$. We begin by noting that even for a large sample size the preventive estimator had the poorest performance. The parametric bootstrap estimator had, in general, a better performance than the MLE, with respect to both bias and MSE. The nonparametric bootstrap estimator showed the same behavior it did for the sample sizes $n=20$ and $n=40$, i.e., had better perfomance for $\theta$ and worse performance for $\beta$ when compared to the MLE. Finally, the best performance was obtained by the corrective estimator, which had, in general, the smallest bias and MSE. Hence, the corrective method can be effective even for large sample sizes.

Tables \ref{tabelasimulmu31} and \ref{tabelasimulmu32} present the results for sample size $n=60$ with respect to the parameter $\mu$. We note that both nonparametric estimators had the worst performance with respect to MSE. Further, the corrective method presented the best performance with respect to MSE, nevertheless it was not much better than the MLE. The parametric bootstrap had a performance worse than both maximum likelihood and the corrective estimator, but better than the nonparametric estimator. But, overall, all estimators were similar. 

Tables \ref{tabelasimulphi31} and \ref{tabelasimulphi32} present the results for sample size $n=60$ with respect to the parameter $\phi$. The MLE had the worst performance with respect to both bias and MSE. Considering the MSE, the best estimator was the parametric bootstrap, which had consistently the smallest MSE. The parametric bootstrap was closely followed by the corrective estimator which had a similar performance, except when the true value of $\phi$ was large, in this case, the MSE of the parametric bootstrap was considerably smaller than the MSE of the corrective method, but, on the other hand, the bias of the corrective method was also considerably smaller than the bias of the parametric bootstrap. Finally, the nonparametric bootstrap outperformed the MLE. Therefore, we conclude that even for sample size large the corrective method and the parametric bootstrap may produce improved estimators.

We have performed a second set of simulations, but now we have two goals. First, to check the performance of the MLE against the proposed estimators, but secondly, we will consider a model which is linearizable, and we will compare the behavior of the estimators for both the linear and non-linear fits. We hope this will give another motivation for the usage of the class of nonlinear regression models that we are introducing here in this paper. This strategy was also used by Cook et al. (1986). In this experiment we consider a linearizable nonlinear model with logit link for the regression parameters and a linearizable nonlinear model with identity link for the precision
parameter:
$${\rm logit}\mu_i = \beta_0x_{i}^{\beta_1},$$
$$\phi_i = \theta_0  x_{i}^{\theta_1},\quad i=1,\ldots,n,$$ 
where the true values of the parameters were taken as $\beta_0 = 0.7$ and $\beta_1 = 0.5$;
and $\theta_0 = 100$ and $\theta_1 = 2$. Note also that here
the elements of the $n\times 2$ matrix $\tilde{X}$ are: $\tilde{X}(\beta)_{i,1} = x_{i}^{\beta_1}$ and $\tilde{X}(\beta)_{i,2} = \beta_0\log(x_{i})(x_{i}^{\beta_1})$. The explanatory variable $x$ was generated
from random draws of an exponential distributions with mean $1$ for $n=20,40$ and $60$. The values of $x$ were held constant throughout the simulations and were the same for both the linearizable and the nonlinear model.
The total number of Monte Carlo replications was set at $5,000$ for each sample size. The number of bootstrap replications was set to 600 for both bootstrap methods.

To fit the linearized model, we consider the model in the following sense:
$$g(\mu_i) = \gamma_0 + \beta_1 z_{i},$$
$$h(\phi_i)= \zeta_0 + \theta_1 z_{i},\quad i=1,\ldots,n,$$
where $\gamma_0 = \log(\beta_0)$, $\zeta_0 = \log(\theta_0)$, $z_{i} = \log(x_i)$, $g(\cdot) = \log({\rm logit}(\cdot))$
and finally, $h(\cdot) = \log(\cdot)$. Hence, we can obtain the MLE of $\beta_0$ in the linearized model simply by exponentiation of $\gamma_0$, and analogously to obtain the MLE of $\theta_0$. 

\begin{table}[t!]
\caption{Simulation results for $n=20$.}
\begin{center}
\begin{tabular}{clccccc}
\hline
Model&Parameter & MLE& Cox-Snell & Firth & p-boot & np-boot\\
\hline
Linear&$\beta_0$ &  0.7008&  0.7020&  0.6998&  0.7022&  0.7040\\
&Bias            &  0.0008&  0.0020& -0.0002&  0.0022&  0.0040\\ 
&Variance        &0.0039&0.0039&0.0039&0.0039&0.0040\\
&MSE             &0.0039&0.0039&0.0039&0.0039&0.0040\\\\
&$\beta_1$       & 0.5025& 0.5014& 0.5035& 0.5014& 0.4991        \\
&Bias            & 0.0025& 0.0014& 0.0035& 0.0014&-0.0009\\
&Variance        &0.0115&0.0115&0.0116&0.0115&0.0132\\
&MSE             &0.0116&0.0115&0.0116&0.0115&0.0132\\\\
&$\theta_0$      &139.37&111.43&171.78&107.73&109.10\\
&Bias            & 39.370& 11.430& 71.780&  7.7300&  9.1000\\
&Variance        &3849.4&2444.3&5917.0&2307.3&2433.6\\
&MSE             &5399.1&2575.0&11070 &2367.1&2516.3\\\\
&$\theta_1$      & 2.0871& 2.0215& 2.1603& 1.9945& 1.9791\\
&Bias            & 0.0871& 0.0215& 0.1603&-0.0055&-0.0209\\
&Variance        &0.1548&0.1489&0.1630&0.1488&0.1530\\
&MSE             &0.1624&0.1494&0.1887&0.1488&0.1535\\\\
Nonlinear&$\beta_0$ &  0.7016&  0.7006&  0.7028&  0.7005&  0.7014\\
&Bias               &  0.0016&  0.0006&  0.0028&  0.0005&  0.0014\\ 
&Variance           &0.0038&0.0038&0.0038&0.0038&0.00394\\
&MSE                &0.0038&0.0038&0.0038&0.0038&0.0039\\\\
&$\beta_1$& 0.5007& 0.4997& 0.5008& 0.4998& 0.4977        \\
&Bias     & 0.0007&-0.0003& 0.0008&-0.0002&-0.0023\\
&Variance &0.0115&0.0115&0.0115&0.0115&0.0128\\
&MSE      &0.0115&0.0115&0.0115&0.0115&0.0129\\\\
&$\theta_0$&140.33&101.14& 199.46& 83.136& 79.011\\
&Bias      & 40.330&  1.1400&  99.460&-16.864&-20.989\\
&Variance  &3822.6&1961.5& 7906.2& 1310.9& 3575.2 \\
&MSE       &5448.8&1962.8& 17798 & 1595.3& 4015.7 \\\\
&$\theta_1$& 2.0923& 2.0264& 2.2071& 1.9991& 1.9863\\
&Bias      & 0.0923& 0.0264& 0.2071&-0.0009&-0.0137\\
&Variance  &0.1558&0.1496&0.1667&0.1482&0.1536\\
&MSE       &0.1643&0.1503&0.2096&0.1482&0.1538\\\\
\hline
\end{tabular}
\end{center}
\label{resulsimullnl}
\end{table}

\begin{table}[t!]
\caption{Simulation results for $n=40$.}
\begin{center}
\begin{tabular}{clccccc}
\hline
Model&Parameter & MLE& Cox-Snell & Firth & p-boot & np-boot\\
\hline
Linear&$\beta_0$ &  0.7001 &  0.7013 &  0.6990 &  0.7018 &  0.7015\\
&Bias            &  0.0001 &  0.0013 & -0.0010 &  0.0018 &  0.0015\\ 
&Variance        &0.0023   &0.0023   &0.0023   &0.0023   &0.0023\\
&MSE             &0.0023   &0.0023   &0.0023   &0.0023   &0.0023\\\\
&$\beta_1$       &  0.5015 & 0.4999  & 0.5030  &  0.4994 &  0.4996       \\
&Bias            &  0.0015 &-0.0001  & 0.0030  & -0.0006 & -0.0004\\
&Variance        &0.0097   &0.0096   &0.0097   &0.0097   &0.0098\\
&MSE             &0.0097   &0.0096   &0.0097   &0.0097   &0.0098 \\\\
&$\theta_0$      &116.87   &  104.55 &  130.11 &  104.18 &  104.25  \\
&Bias            &16.870   &  4.5500 &  30.110 &  4.1800 & 4.2500  \\
&Variance        &1081.2   &  861.77 &  1348.1 &  882.60 &  862.99  \\
&MSE             &1365.7   &  882.46 &  2254.8 &  900.09 &  881.06  \\\\
&$\theta_1$      &  2.0346 &  2.0076 &  2.0627 &  2.0029 &   1.9939 \\
&Bias            &  0.0346 &  0.0076 &  0.0627 &  0.0029 &  -0.0061 \\
&Variance        &0.0663   &0.0647   &0.0682   &0.0662   & 0.0657 \\
&MSE             &0.0675   &0.0647   &0.0721   &0.0662   & 0.0657 \\\\
Nonlinear&$\beta_0$ &  0.7002 &  0.6999 &  0.7005   &  0.6999 &  0.7000\\
&Bias               &  0.0002 & -0.0001 &  0.0005   & -0.0001 &  0.0000\\ 
&Variance           &0.0023   &0.0023   &0.0023     &0.0023   &0.0023\\
&MSE                &0.0023   &0.0023   &0.0023     &0.0023   &0.0023\\\\
&$\beta_1$          &  0.5017 &  0.5001 &  0.5029   &  0.5002 &  0.4998        \\
&Bias               &  0.0017 &  0.0001 &  0.0029   &  0.0002 & -0.0002\\
&Variance           &0.0096   &0.0095   &0.0096     &0.0095   &0.0097\\
&MSE                &0.0096   &0.0095   &0.0096     &0.0095   &0.0097\\\\
&$\theta_0$         & 116.85  &   100.38&  137.00   &   97.198& 97.435\\
&Bias               &  16.850 &   0.3800&  37.000   &  -2.8020& -2.5650\\
&Variance           & 1057.6  &   775.37&  1466.5   &   724.96&   766.40 \\
&MSE                & 1341.7  &   775.51&  2835.2   &   732.81&   772.98 \\\\
&$\theta_1$         &  2.0327 &  2.0054 &  2.0726   &   1.9997&   1.9929\\
&Bias               &  0.0327 &  0.0054 & 0.0726    &  -0.0003&  -0.0071\\
&Variance           &0.0631   &0.0616   & 0.0649    & 0.0617  & 0.0627  \\
&MSE                &0.0642   &0.0616   & 0.0702    & 0.0617  & 0.0627
\\\\
\hline
\end{tabular}
\end{center}
\label{resulsimullnl2}
\end{table}

\begin{table}[t!]
\caption{Simulation results for $n=60$.}
\begin{center}
\begin{tabular}{clccccc}
\hline
Model&Parameter & MLE& Cox-Snell & Firth & p-boot & np-boot\\
\hline
Linear&$\beta_0$ &   0.6718  &  0.6768   & 0.6712 & 0.7322  & 0.6659\\
&Bias            &  -0.0282  & -0.3232   &-0.0288 & 0.0322  &-0.0341\\ 
&Variance        &  0.0101   &0.0076     &0.0101  &0.0109   &0.0146\\
&MSE             &  0.0109   &0.0082     &0.0109  &0.0119   &0.0158\\\\
&$\beta_1$       &  0.5435   &  0.5327   & 0.5442 & 0.4705  & 0.5582\\
&Bias            &  0.0435   &  0.0327   & 0.0442 &-0.0295  & 0.0582\\
&Variance        & 0.0249    & 0.0168    &0.0249  &0.0200   &0.0422\\
&MSE             & 0.0268    & 0.0178    &0.0268  &0.0209   &0.0455\\\\
&$\theta_0$      &    100.99 &   93.896  &  108.46& 108.93  & 93.676  \\
&Bias            &    0.9900 &   6.1040  &  8.4600& 8.9300  & -6.3240  \\
&Variance        &    1471.2 &   1270.5  &  1703.6& 1307.1  & 1299.8  \\
&MSE             &    1472.2 &   1307.7  &  1775.2& 1386.8  & 1339.8  \\\\
&$\theta_1$      &    1.9216 &  1.9064   &  1.9358& 2.0261  & 1.8697  \\
&Bias            &   -0.0784 & -0.0936   & -0.0642&   0.0261&-0.1303  \\
&Variance        &   0.1029  & 0.1032    & 0.1059 &0.1629   &0.1693  \\
&MSE             &   0.1090  & 0.1120    & 0.1100 &0.1636   &0.1862\\\\
Nonlinear&$\beta_0$ &  0.6977&   0.6975  &  0.6981&  0.6976 &  0.6973 \\
&Bias               & -0.0023&  -0.0025  & -0.0019& -0.0024 & -0.0027 \\ 
&Variance           &0.0033  & 0.0033    &0.0033  &0.0033   &0.0035 \\
&MSE                &0.0033  & 0.0033    &0.0033  &0.0033   &0.0035 \\\\
&$\beta_1$          &  0.4818&   0.4925  &  0.4716&  0.4826 &  0.4862         \\
&Bias               & -0.0182&  -0.0075  & -0.0284& -0.0174 & -0.0138 \\
&Variance           &  0.0212& 0.0058    &  0.0415&  0.0201 &  0.01867 \\
&MSE                &  0.0212& 0.0059    &  0.0416&  0.0201 &  0.01869 \\\\
&$\theta_0$         &   109.19&  100.93   &   120.89 &  99.256 &   97.975 \\
&Bias               &   9.1900&    0.9300 &    20.890&  -0.7440&   -2.0250 \\
&Variance           &   635.41&  451.46   &   777.70 &  532.02 &   522.11  \\
&MSE                &   719.92&  452.33   &   1214.2 &  532.57 &   526.21  \\\\
&$\theta_1$         &   1.9989&   1.9839  &  2.0256  &  1.9936 &    1.9776\\
&Bias               &  -0.0011&  -0.0161  &  0.0256  & -0.0064 &   -0.0224\\
&Variance           & 0.0323  & 0.0320    &0.0373    &0.0333   &  0.0317  \\
&MSE                & 0.0323  & 0.0323    &0.0380    &0.0333   &  0.0322 \\\\
\hline
\end{tabular}
\end{center}
\label{resulsimullnl3}
\end{table}

Table \ref{resulsimullnl} presents the results for $n=20$. The best estimator, with respect to both bias and MSE, was the parametric bootstrap. The preventive estimator had the poorest performance. Considering both the linear and nonlinear models, the nonparametric bootstrap had better performance than the MLE for the parameter $\theta$ and worse for the parameter $\beta$. The corrective estimator had a good performance, in the sense that only the parametric bootstrap had outperformed it. Now, comparing the linear and nonlinear models, we note that the MLE was similar for both models. Comparing the corrective and parametric bootstrap estimators for both models, the estimators from the nonlinear model had the best performance. The nonparametric bootstrap estimator had a similar performance in both models, except for $\theta_0$ which was considerably worse in the nonlinear model. Hence, we conclude that for the estimators of interest, i.e., corrective and parametric bootstrap, the nonlinear model had the best performance and should be used instead of its linearized version if one aims to obtain better performance with respect to MSE and bias.

Table \ref{resulsimullnl2} presents the results for $n=40$. The best performance, with respect to both bias and MSE, was achieved by the corrective estimator. Considering both the linear and nonlinear models, the nonparametric bootstrap had better performance than the MLE for the parameter $\theta$ and worse for the parameter $\beta$. The preventive estimator had the worst performance. The parametric bootstrap estimator had a good performance, in the sense that only the corrective estimator had a best performance. Now, comparing the linear and nonlinear models, we note that the MLE was similar for both models. Comparing the corrective, parametric bootstrap and nonparametric estimators for both models, the estimators from the nonlinear model had the best performance. Here, we conclude that, for this sample size, the corrective estimator along with the bootstrap based estimators had a better performance when modelling the nonlinear model, than when modelling its linearized version.

Table \ref{resulsimullnl3} presents the results for $n=60$. The best performance, with respect to both bias and MSE, was achieved by the corrective and parametric bootstrap estimators, however, their estimates were not far from the ones obtained by the MLE, this is probably due to the large sample size. The preventive estimator had the worst performance and the nonparametric bootstrap estimator had a similar performance to the MLE. It is noteworthy  that the best estimates were obtained by the corrective estimator in the nonlinear model, followed by the parametric bootstrap estimator also in the nonlinear model. Comparing the linear and nonlinear models, we note that the MLE was similar for both models. Comparing the corrective, parametric bootstrap and nonparametric estimators for both models, the estimators from the nonlinear model had the best performance. Therefore, we conclude that for all sample sizes the nonlinear model should be preferred to its linearized version. 

\section{Application to real data}
We now present an application of the linear beta regression model with dispersion covariates, the particular case considered in Section \ref{partcases}.
The source of the data is Prater (1956). We want to model the proportion of crude oil that is converted to petrol after fractional distillation. We have
two explanatory variables for this model. The first is the level of crude oil, where $10$ different possible levels correspond to the proportion of
crude oil that was vaporized. The second is the temperature in Fahrenheit at which all petrol that is contained in the amount of crude oil vaporizes.
Ferrari and Cribari-Neto (2004) and Ospina et al. (2006) used this data as an illustration of the linear beta
regression model and of some bias-correction schemes for the model, respectively.

The sample size is $n=32$. The model specification consists of two parts as seen in equation (\ref{regr}). The first, which is related to the mean,
includes an intercept $(x_1 = 1)$, $9$ different dummy variables $(x_2,\ldots,x_{10})$ to represent the $10$ possible different situations for the level
of crude oil and the covariate $x_{11}$, measuring the temperature in Fahrenheit degrees at which all petrol vaporizes. The second, which is related
to the precision parameter, includes an intercept $(z_1 = 1)$ and the covariate $z_2 = x_{11}$.

\begin{table}[htb]
\caption{Statistics and $p$-values of the LRT and ST.}
\begin{center}
\begin{tabular}{ccc}
\hline
Test & Stat. value & $p$-value\\
\hline
LRT & $4.35902$ & $0.03681$ \\
ST& $6.57124$ & $0.01036$ \\
\hline
\end{tabular}
\end{center}
\label{tabelapvaloraplic}
\end{table}

\begin{table}[htb]
\caption{Estimated values of the parameters and their standard errors.}
\begin{center}
\begin{tabular}{cccccc}
\hline
Parameter & MLE& Cox-Snell & Firth & p-boot & np-boot\\
\hline
$\beta_1$ &  -5.92323  &-5.91695&-5.92336       &-5.91213  &-5.92425   \\
          &   (0.18352)&(0.22155) &(0.15302)    & (0.20352)& (0.14326) \\
$\beta_2$&    1.60198  & 1.60063&1.60179        & 1.60231  & 1.61022  \\
         &    (0.06385)&(0.08602) &(0.05271)    & (0.06851)& (0.07535)\\
$\beta_3$&    1.29726  & 1.29592&1.29800        & 1.29429  & 1.29766   \\
         &   (0.09910) &(0.12392)  &(0.08239)   &(0.10315) &(0.09109) \\
$\beta_4$&   1.56533   & 1.56363&1.56472        &1.56333   &1.58932    \\
         &    (0.09973)&(0.12376)&(0.08300)     & (0.11973)& (0.12473) \\
$\beta_5$&    1.03007  & 1.02919&1.03019        & 1.03334  & 1.03327   \\
         &    (0.06328)& (0.08554)&(0.05223)    & (0.06728)& (0.06998) \\
$\beta_6$&    1.15416  & 1.15318&1.15416        & 1.15399  & 1.15456   \\
         &   (0.06564) &(0.08771)&(0.05424)     &(0.07533) &(0.06167)  \\
$\beta_7$&    1.01944  & 1.01857&1.01991        & 1.01744  & 1.01934   \\
         &   (0.06635) & (0.08925)&(0.05478)    &(0.06323) &(0.06835)  \\
$\beta_8$&    0.62225  & 0.62171&0.62038        & 0.62111  & 0.62448   \\
         &   (0.06563) &(0.08921)&(0.05416)     &(0.07563) &(0.07599)  \\
$\beta_9$&    0.56458  & 0.56415&0.56485        & 0.56558  & 0.56651   \\
         &    (0.06018)&(0.08321)&(0.04957)     & (0.07066)& (0.07538) \\
$\beta_{10}$&   0.35943& 0.35906&0.36046        &  0.35911&   0.36433 \\
         &   (0.06714) &(0.09243) &(0.05532)    &(0.08332) &(0.07755)  \\
$\beta_{11}$&   0.01035& 0.01034&0.01035        &   0.01035&   0.01037 \\
         &   (0.00043) & (0.00052)&(0.00036)    &(0.00049) &(0.00066)  \\
$\theta_1$&   1.36408  & 1.98699&1.64216        & 1.88568  & 1.45548   \\
         &   (1.22578) &(1.22669)&(1.22819)     &(1.22877) &(1.22439)  \\
$\theta_2$&   0.01457  & 0.01147&0.01483        & 0.01266 & 0.01387    \\
          &   (0.00361)& (0.00362)&(0.00362)    & (0.00374)& (0.00369)\\
\hline
\end{tabular}
\end{center}
\label{tabelaaplic}
\end{table}

\begin{table}[htb]
\caption{Estimated values of $\mu$ and $\phi$.}
\begin{center}
\begin{tabular}{ccccccccccc}
\hline
$i$ & \multicolumn{2}{c}{MLE}& \multicolumn{2}{c}{Cox-Snell} & \multicolumn{2}{c}{Firth} & \multicolumn{2}{c}{p-boot} & \multicolumn{2}{c}{np-boot}\\
    & $\mu_i$        &$\phi_i$& $\mu_i$        &$\phi_i$& $\mu_i$        &$\phi_i$& $\mu_i$        &$\phi_i$& $\mu_i$        &$\phi_i$\\
    \hline
 $1$ &0.0999&  77.500&0.0999& 66.000&0.0996&  92.800&0.0996&  66.900&0.0998&  53.600 \\
 $2$ &0.1865& 215.00&0.1866&154.50&0.1863& 279.50&0.1863& 203.80&0.1863& 200.00 \\
 $3$ &0.3214& 596.30&0.3214&311.10&0.3212& 798.10&0.3213& 577.10&0.3214& 658.30 \\
 $4$ &0.4737&1471.7&0.4737&432.20&0.4736&1941.1&0.4737&1371.2&0.4737&1772.5 \\
 $5$ &0.0856&  93.700&0.0855& 77.900&0.0854& 114.40&0.0853&  82.800&0.0853&  69.400 \\
 $6$ &0.1421& 208.80&0.1419&151.10&0.1418& 271.00&0.1418& 197.60&0.1419& 193.00 \\
 $7$ &0.2628& 613.90&0.2626&316.30&0.2625& 821.80&0.2625& 593.90&0.2625& 680.20 \\
 $8$ &0.1032&  85.800&0.1031& 72.200&0.1028& 103.90&0.1029&  75.100&0.1029&  61.700 \\
 $9$ &0.1765& 205.80&0.1763&149.40&0.1760& 266.90&0.1761& 194.60&0.1761& 189.60 \\
$10$ &0.3024& 554.40&0.3022&298.20&0.3019& 741.60&0.3021& 536.90&0.3023& 606.30 \\
$11$ &0.0788& 120.00&0.0787& 96.300&0.0785& 149.90&0.0786& 108.90&0.0786&  96.200 \\
$12$ &0.1436& 309.50&0.1437&203.00&0.1434& 408.80&0.1435& 297.80&0.1435& 309.40 \\
$13$ &0.2475& 798.10&0.2475&362.10&0.2473&1067.5&0.2474& 767.40&0.2475& 911.30 \\
$14$ &0.3439&1537.5&0.3439&432.60&0.3437&2024.5&0.3438&1427.9&0.3437&1857.2 \\
$15$ &0.1695& 342.80&0.1696&218.20&0.1693& 454.30&0.1694& 330.70&0.1694& 348.80 \\
$16$ &0.2754& 821.70&0.2755&366.90&0.2752&1098.8&0.2753& 789.40&0.2753& 941.10 \\
$17$ &0.3369&1235.6&0.3369&422.30&0.3367&1639.2&0.3368&1164.7&0.3368&1469.0 \\
$18$ &0.1054& 191.30&0.1054&141.00&0.1053& 247.10&0.1052& 180.10&0.1052& 173.40 \\
$19$ &0.2360& 742.00&0.2361&349.60&0.2359& 993.00&0.2359& 715.00&0.2359& 840.60 \\
$20$ &0.3231&1368.3&0.3231&429.60&0.3230&1809.4&0.3230&1281.4&0.3230&1639.4 \\
$21$ &0.0538& 120.00&0.0538& 96.300&0.0535& 149.90&0.0536& 108.90&0.0536&  96.200 \\
$22$ &0.0792& 215.00&0.0792&154.50&0.0789& 279.50&0.0791& 203.80&0.0791& 200.00 \\
$23$ &0.1690& 720.70&0.1691&344.50&0.1686& 964.60&0.1689& 695.00&0.1689& 813.80 \\
$24$ &0.2706&1677.9&0.2706&430.20&0.2700&2201.4&0.2705&1547.6&0.2704&2038.3 \\
$25$ &0.0827& 248.70&0.0827&172.80&0.0825& 325.70&0.0825& 237.50&0.0825& 238.50 \\
$26$ &0.1711& 798.10&0.1712&362.10&0.1710&1067.50&0.1710& 767.40&0.1711& 911.30 \\
$27$ &0.3188&2523.2&0.3188&337.10&0.3187&3238.4&0.3187&2236.3&0.3187&3129.6 \\
$28$ &0.1270& 650.80&0.1270&326.60&0.1269& 871.20&0.1268& 628.90&0.1268& 726.20 \\
$29$ &0.2366&1885.4&0.2366&419.20&0.2365&2460.2&0.2364&1721.5&0.2365&2306.0 \\
$30$ &0.1050& 798.10&0.1051&362.10&0.1049&1067.5&0.1049& 767.40&0.1049& 911.30 \\
$31$ &0.1195& 978.70&0.1195&394.30&0.1193&1305.5&0.1194& 933.80&0.1195&1140.4 \\
$32$ &0.1840&1998.5&0.1840&409.80&0.1838&2600.1&0.1839&1814.9&0.1840&2452.1 \\

\hline
\end{tabular}
\end{center}
\label{tabelaaplic2}
\end{table}

The logit link function was used to relate the mean of the response variable to the linear predictor, and the log link function was used
to relate the precision parameter to its linear predictor. The unknown coefficients were estimated through maximum likelihood using the quasi-Newton
optimization method BFGS (see, for instance, Press et al., 1992) with analytical derivatives. The corrective (based on Cox and Snell, 1968),
preventive (based on Firth, 1993) and bootstrap bias corrected (based on Efron, 1979) bias corrected schemes considered in Sections \ref{biasbetatheta} and \ref{biasmuphi}
were also computed.

We can then write the model we are considering as
$${\rm logit}(\mu_i) = \beta_1 + \beta_2 x_{2i} + \beta_3 x_{3i} + \beta_4 x_{4i}+ \beta_5 x_{5i}+ \beta_6 x_{6i}+ \beta_7 x_{7i}+ \beta_8 x_{8i}+ \beta_9 x_{9i}+ \beta_{10} x_{10i}+ \beta_{11} x_{11i},$$
$$\log(\phi_i) = \theta_1 + \theta_2 z_{2i},\quad i=1,\ldots,32,$$
where ${\rm logit}(x) = \log(x/(1-x))$.

Looking at Table \ref{tabelaaplic} we see that $\theta_2$ has an standard error of $0.00361$. Thus, if we use the asymptotic normality to test if $\theta_2=0$,
one would obtain the $p$-value $5.65\times 10^{-5}$, which indicates the significance of this parameter. As a further study on this direction we considered the
likelihood ratio test (LRT) and the score test (ST), to test the null hypothesis that the true model does not contain $\theta_2$ (that is, $\theta_2=0$),
which was the model considered by Ferrari and Cribari-Neto (2004), and Ospina et al. (2006), versus, the full model considered above (that is, $\theta_2\neq 0$).
The value of the statistics together with the $p$-value obtained by the comparison with the quantiles of a $\chi_1^2$ distribution are given below in Table \ref{tabelapvaloraplic}.
By looking at Table \ref{tabelapvaloraplic} we may conclude that the null hypothesis (that $\theta_2=0$) should be rejected, thus showing that the
general model introduced in Section 2 may be useful to practioners.

Table \ref{tabelaaplic} presents the maximum likelihood estimates along with their corrected versions and the corresponding standard errors. The
adjusted versions are the corrective bias-corrected estimator (Cox-Snell), the preventive bias-corrected estimator (Firth), the parametric
bootstrap bias-corrective estimator (p-boot) and the nonparametric bootstrap bias-corrected estimator (np-boot). It can be
seen that the estimates and corrected estimates for the parameters $\beta$'s are very similar, however, for the parameters $\theta$'s, we observe
some difference between the maximum likelihood estimate and the corrected ones. 

In Table \ref{tabelaaplic2} we give the maximum likelihood estimates of $\mu$ and $\phi$ together with their corrected versions. We then see that
there are not large differences between the estimates of $\mu$ and the corrected estimates of $\mu$, nevertheless, for the parameter $\phi$, we note
considerably differences between the maximum likelihood estimates and the corrected ones.

\section{Conclusions}
We defined a general beta regression models which allows a regression structure on the
precision parameter, and both the regression structures on the mean and on the precision
parameters are allowed to be nonlinear. Then, using the approximation theory developed
by Cox and Snell (1968), we
calculate ${\cal O}\left (n^{-1}\right )$ bias for the MLEs for $\beta$ and $\theta$. 
Our results generalize the formulae obtained by Ospina et al. (2006). We then defined bias-free estimator
to order ${\cal O}\left (n^{-1}\right )$, by using the expressions obtained through Cox and Snell's (1968)
formulae and Firth's (1993) estimating equation. We also considered two schemes of bias correction based
on bootstrap. We use simulation in a nonlinear beta regression model with nonlinear dispersion covariates
to conclude the superiority of the corrective and parametric bootstrap methods of bias correction over the other methods presented here with regard to both bias reduction and mean squared error. In fact, the parametric bootstrap presented, in general, the smallest mean squared error, and the corrective method the smallest bias. Further, the corrective method has an advantage over the parametric bootstrap, which is the fact that the parametric bootstrap is computer intensive whereas the corrective method is not. The simulation also considered another study, to check if a nonlinear model which can be linearized should be estimated through a nonlinear model, or through a linear model. The simulation showed that the nonlinear model should be preferred, showing that the general model we are presenting has the potential to be very useful to practioners. The paper is concluded with an empirical application
to illustrate the usefulness of our results, and more speciffically, the usefulness of considering a model with dispersion covariates.

\section*{Acknowledgements}

The first author is grateful to CNPq for their finantial support, the second and third authors are grateful to CAPES for their financial support.
\section*{Appendix}

We give explicit expressions for the cumulants and their derivatives, both defined in page \pageref{cumulants}. Further, we give the expressions for each
quantity contained in equation (\ref{coxsnell}), some of them are also deduced to help the reader who might be interested in checking the results.

We call attention for the fact that the expressions of the type contained in equation (\ref{coxsnell}) from the Appendix of Ospina et al. (2006)
are incorrect if we put $a$ as $\phi$. But this is a minor issue, and the formulae contained in the body of the text are correct.

Consider initially the following notation for the derivatives, and product of the derivatives, of the predictor with respect to the regression parameters:
$$(rs)_i = \frac{\partial^2\eta_{1i}}{\partial\beta_r\partial\beta_s},\quad (RS)_i = \frac{\partial^2\eta_{2i}}{\partial\theta_R\partial\theta_S},\quad (rs,T)_i = \frac{\partial^2\eta_{1i}}{\partial\beta_r\partial\beta_s}\frac{\partial\eta_{2i}}{\partial\theta_T},$$
and so on. Now, consider the following quantities:
\begin{eqnarray*}
a_i &=& \psi'((1-\mu_i)\phi_i) + \psi'(\mu_i\phi_i),\\
b_i &=& \psi'((1-\mu_i)\phi_i)(1-\mu_i)^2 + \psi'(\mu_i\phi_i)\mu_i^2 - \psi'(\phi_i),\\
c_i &=& \psi''((1-\mu_i)\phi_i) - \psi''(\mu_i\phi_i),\\
d_i &=& \psi''((1-\mu_i)\phi_i)(1-\mu_i)^2 - \psi''(\mu_i\phi_i)\mu_i^2,\\
e_i &=& \psi''(\phi_i) - \psi''(\mu_i\phi_i)\mu_i^3 - \psi''((1-\mu_i)\phi_i)(1-\mu_i)^3.
\end{eqnarray*}

By using these quantities, the cumulants can be written as
\begin{eqnarray*}
\kappa_{rs} &=& -\sum_{i=1}^n \phi_i^2 a_i \left(\frac{d\mu_i}{d\eta_{1i}}\right)^2 (r,s)_i,\\
\kappa_{rS} &=& -\sum_{i=1}^n \phi_i \{ \mu_i a_i -\psi'((1-\mu_i)\phi_i)\}\frac{d\mu_i}{d\eta_{1i}}\frac{d\phi_i}{d\eta_{2i}}(r,S)_i,\\
\kappa_{RS} &=& -\sum_{i=1}^n b_i \left(\frac{d\phi_i}{d\eta_{2i}}\right)^2 (R,S)_i,
\end{eqnarray*}
\begin{eqnarray*}
\kappa_{rsu} &=& \sum_{i=1}^n\phi_i^2\left\{ \phi_i c_i \left(\frac{d\mu_i}{d\eta_{1i}}\right)^3 - 3 a_i \frac{d\mu_i}{d\eta_{1i}}\frac{d^2\mu_i}{d\eta_{1i}^2}\right\} (r,s,u)_i\\
&&- \sum_{i=1}^n \phi_i^2 a_i \left( \frac{d\mu_i}{d\eta_{1i}}\right)^2 \{ (rs,u)_i + (ru,s)_i + (su,r)_i\},\\
\kappa_{rsU} &=& -\sum_{i=1}^n \phi_i\{2a_i + \phi_i \psi''((1-\mu_i)\phi_i) - \phi_i\mu_i c_i\}\left(\frac{d\mu_i}{d\eta_{1i}}\right)^2\frac{d\phi_i}{d\eta_{2i}}(r,s,U)_i\\
&&+\sum_{i=1}^n\phi_i\{\psi'((1-\mu_i)\phi_i)-\mu_i a_i\}\frac{d^2\mu_i}{d\eta_{1i}^2}\frac{d\phi_i}{d\eta_{2i}}(r,s,U)_i\\
&&+ \sum_{i=1}^n \phi_i \{ \psi'((1-\mu_i)\phi_i) -\mu_i a_i\} \frac{d\mu_i}{d\eta_{1i}}\frac{d\phi_i}{d\eta_{2i}} (rs,U)_i,
\end{eqnarray*}
\begin{eqnarray*}
\kappa_{rSU} &=&\sum_{i=1}^n \{ \psi'((1-\mu_i)\phi_i) - \mu_i a_i\} \left\{ 2 \frac{d\mu_i}{d\eta_{1i}}\left(\frac{d\phi_i}{d\eta_{2i}}\right)^2+\phi_i\frac{d\mu_i}{d\eta_{1i}}\frac{d^2\phi_i}{d\eta_{2i}^2}\right\}(r,S,U)_i\\
&&+ \sum_{i=1}^n \phi_i d_i \frac{d\mu_i}{d\eta_{1i}}\left(\frac{d\phi_i}{d\eta_{2i}}\right)^2 (r,S,U)_i\\
&&+\sum_{i=1}^n \phi_i \{\psi'((1-\mu_i)\phi_i) - \mu_i a_i\} \frac{d\mu_i}{d\eta_{1i}}\frac{d\phi_i}{d\eta_{2i}} (r,SU)_i,\\
\kappa_{RSU} &=& \sum_{i=1}^n \left\{ e_i \left(\frac{d\phi_i}{d\eta_{2i}}\right)^3 - 3 b_i \frac{d^2\phi_i}{d\eta_{2i}^2}\frac{d\phi_i}{d\eta_{2i}}\right\}(R,S,U)_i\\
&&-\sum_{i=1}^n b_i \left(\frac{d\phi_i}{d\eta_{2i}}\right)^2 \{ (RS,U)_i + (RU,S)_i + (SU,R)_i\}.
\end{eqnarray*}

Differentiating the second order cumulants with respect to the parameters, we have
\begin{eqnarray*}
\kappa_{rs}^{(u)} &=& - \sum_{i=1}^n \phi_i^2 \left\{2a_i \frac{d\mu_i}{d\eta_{1i}} \frac{d^2\mu_i}{d\eta_{1i}^2} - \phi_i c_i \left(\frac{d\mu_i}{d\eta_{1i}}\right)^3\right\} (r,s,u)_i\\
&&- \sum_{i=1}^n \phi_i^2 a_i \left(\frac{d\mu_i}{d\eta_{1i}}\right)^2 \{(ru,s)_i + (su,r)_i\},\\
\kappa_{rs}^{(U)} &=& - \sum_{i=1}^n \left\{ \phi_i^2 [ \psi''((1-\mu_i)\phi_i) - \mu_i c_i] + 2\phi_i a_i\right\} \left(\frac{d\mu_i}{d\eta_{1i}}\right)^2 \frac{d\phi_i}{d\eta_{2i}} (r,s,U)_i,\\
\kappa_{RS}^{(u)} &=& \sum_{i=1}^n  \{ d_i\phi_i + 2 \psi'((1-\mu_i)\phi_i) - 2\mu_i a_i\} \frac{d\mu_i}{d\eta_{1i}} \left(\frac{d\phi_i}{d\eta_{2i}}\right)^2 (R,S,u)_i,\\
\kappa_{RS}^{(U)} &=& \sum_{i=1}^n \left\{ e_i \left(\frac{d\phi_i}{d\eta_{2i}} \right)^3 - 2 b_i \frac{d\phi_i}{d\eta_{2i}} \frac{d^2\phi_i}{d\eta_{2i}^2}\right\}(R,S,U)_i\\
&&- \sum_{i=1}^n b_i \left(\frac{d\phi_i}{\eta_{2i}} \right)^2 \{ (RU,S)_i + (SU,R)_i \},
\end{eqnarray*}
\begin{eqnarray*}
\kappa_{rS}^{(u)} &=& \sum_{i=1}^n \phi_i \{ \phi_i\mu_i c_i - \psi''((1-\mu_i)\phi_i)\phi_i - a_i\} \left(\frac{d\mu_i}{d\eta_{1i}} \right)^2 \frac{d\phi_i}{d\eta_{2i}} (r,S,u)_i\\
&&+ \sum_{i=1}^n \phi_i \{ \psi'((1-\mu_i)\phi_i) - a_i \mu_i\}\frac{d^2\mu_i}{d\eta_{1i}^2} \frac{d\phi_i}{d\eta_{2i}} (r,S,u)_i\\
&&+ \sum_{i=1}^n \phi_i \{ \psi'((1-\mu_i)\phi_i) - a_i \mu_i\} \frac{d\mu_i}{d\eta_{1i}}\frac{d\phi_i}{d\eta_{2i}}(ru,S)_i,
\end{eqnarray*}
\begin{eqnarray*}
\kappa_{rS}^{(U)} &=& \sum_{i=1}^n \{ \psi'((1-\mu_i)\phi_i) - a_i \mu_i + \phi_i d_i\} \frac{d\mu_i}{d\eta_{1i}} \left(\frac{d\phi_i}{d\eta_{2i}} \right)^2(r,S,U)_i\\
&&+ \sum_{i=1}^n \phi_i \{\psi'((1-\mu_i)\phi_i) - a_i\mu_i\} \frac{d\mu_i}{d\eta_{1i}} \frac{d^2\phi_i}{d\eta_{2i}^2}(r,S,U)_i\\
&&+ \sum_{i=1}^n \phi_i \{ \psi'((1-\mu_i)\phi_i) - a_i\mu_i\} \frac{d\mu_i}{d\eta_{1i}} \frac{d\phi_i}{d\eta_{2i}} (SU,r)_i.
\end{eqnarray*}
We now define the diagonal matrices
\begin{eqnarray}
M_1 &=& {\rm diag} \left( \frac{\phi_i^2}{2} \left[ \phi_i c_i \left(\frac{d\mu_i}{d\eta_{1i}} \right)^3 - a_i \frac{d\mu_i}{d\eta_{1i}} \frac{d^2\mu_i}{d\eta_{1i}^2}\right]\right),\label{matrizm1}\\
M_2 &=& {\rm diag} \left( \frac{\phi_i^2}{2} \{\mu_i c_i - \psi''((1-\mu_i)\phi_i) \}\left(\frac{d\mu_i}{d\eta_{1i}} \right)^2\frac{d\phi_i}{d\eta_{2i}}\right.\nonumber\\
&&\left.+\phi_i\{\psi'((1-\mu_i)\phi_i) - a_i\mu_i\}  \frac{d^2\mu_i}{d\eta_{1i}^2}\frac{d\phi_i}{d\eta_{2i}} \right)\\
M_3 &=& {\rm diag} \left( -\frac{\phi_i}{2} \left\{[2a_i + \phi_i \psi''((1-\mu_i)\phi_i) - \phi_i \mu_i c_i] \left(\frac{d\mu_i}{d\eta_{1i}} \right)^2 \frac{d\phi_i}{d\eta_{2i}}    \right.  \right.\nonumber\\
&&\left.\left.+ [\psi'((1-\mu_i)\phi_i) - \mu_i a_i] \frac{d^2\mu_i}{d\eta_{1i}^2} \frac{d\phi_i}{d\eta_{2i}} \right\}\right),\\
M_4 &=& {\rm diag} \left( \frac{1}{2} \left\{ [d_i\phi_i + 2\psi'((1-\mu_i)\phi_i) - 2\mu_i a_i] \frac{d\mu_i}{d\eta_{1i}} \left(\frac{d\phi_i}{d\eta_{2i}} \right)^2 \right.\right.\nonumber\\
&&\left.\left. - \phi_i[\psi'((1-\mu_i)\phi_i) - \mu_i a_i] \frac{d\mu_i}{d\eta_{1i}} \frac{d^2\phi_i}{d\eta_{2i}^2}\right\}\right),\\
M_5 &=& {\rm diag} \left( \frac{\phi_i}{2} \left\{ d_i \frac{d\mu_i}{d\eta_{1i}} \left( \frac{d\phi_i}{d\eta_{2i}} \right)^2 + [\psi'((1-\mu_i)\phi_i) -\mu_i a_i]\frac{d\mu_i}{d\eta_{1i}} \frac{d^2\phi_i}{d\eta_{2i}^2}\right\}\right),\\
M_6 &=& {\rm diag} \left( \frac{1}{2} \left[e_i \left(\frac{d\phi_i}{d\eta_{2i}} \right)^3-b_i \frac{d\phi_i}{d\eta_{2i}} \frac{d^2\phi_i}{d\eta_{2i}^2} \right]\right),\label{matrizm6}
\end{eqnarray}
\begin{eqnarray}
N_1 &=& {\rm diag}\left( \frac{\phi_i^2 a_i}{2} \left(\frac{d\mu_i}{d\eta_{1i}} \right)^2\right)\label{matrizn1},\\
N_2 &=& {\rm diag} \left(\frac{\phi_i}{2} [\psi'((1-\mu_i)\phi_i) - a_i\mu_i]\frac{d\mu_i}{d\eta_{1i}}\frac{d\phi_i}{d\eta_{2i}} \right),\label{matrizn2} \\
N_3 &=& {\rm diag} \left(\frac{b_i}{2} \left(\frac{d\phi_i}{d\eta_{2i}} \right)^2 \right)\label{matrizn3}.
\end{eqnarray}

Let now, $m_{ji}$ be the $i$th diagonal element of the matrix $M_j$, and $n_{ji}$ be the $i$th diagonal element of the matrix $N_j$. Thus, consider also
the following matrices
\begin{eqnarray*}
L_1 &=& \sum_{i=1}^n n_{2i} \tilde{X}_i K^{\beta\theta} \tilde{Z}^T \delta_i,\\
L_2 &=& \sum_{i=1}^n n_{2i} \tilde{Z}_i K^{\beta\theta} \tilde{X}^T \delta_i,
\end{eqnarray*}
where $\delta_i$ is an $n\times 1$ vector with a one in the $i$th position and zeros elsewhere.

From expressions (\ref{matrizm1}) to (\ref{matrizn3}) we get:
$$\kappa_{rs}^{(u)} - \frac{1}{2} \kappa_{rsu} = \sum_{i=1}^n m_{1i} (r,s,u)_i + \sum_{i=1}^n n_{1i} \{ (rs,u)_i - (ru,s)_i - (su,r)_i\},$$
$$\kappa_{Rs}^{(u)} - \frac{1}{2} \kappa_{Rsu} = \sum_{i=1}^n m_{2i} (R,s,u)_i + \sum_{i=1}^n n_{2i} (su,R)_i,$$
$$\kappa_{rS}^{(u)} - \frac{1}{2} \kappa_{rSu} = \sum_{i=1}^n m_{2i} (r,S,u)_i + \sum_{i=1}^n n_{2i} (ru,S)_i,$$
$$\kappa_{rs}^{(U)} - \frac{1}{2} \kappa_{rsU} = \sum_{i=1}^n m_{3i} (r,s,U)_i - \sum_{i=1}^n n_{2i} (rs,U)_i,$$
$$\kappa_{RS}^{(u)} - \frac{1}{2} \kappa_{RSu} = \sum_{i=1}^n m_{4i} (R,S,u)_i - \sum_{i=1}^n n_{2i} (RS,u)_i,$$
$$\kappa_{Rs}^{(U)} - \frac{1}{2} \kappa_{RsU} = \sum_{i=1}^n m_{5i} (R,s,U)_i + \sum_{i=1}^n n_{2i} (RU,s)_i,$$
$$\kappa_{rS}^{(U)} - \frac{1}{2} \kappa_{rSU} = \sum_{i=1}^n m_{5i} (r,S,U)_i + \sum_{i=1}^n n_{2i} (SU,r)_i,$$
$$\kappa_{RS}^{(U)} - \frac{1}{2} \kappa_{RSU} = \sum_{i=1}^n m_{6i} (R,S,U)_i + \sum_{i=1}^n n_{3i} \{(RS,U)_i - (RU,S)_i - (SU,R)_i\}.$$

We are now in conditions to compute each term of expression (\ref{coxsnell}):
\begin{eqnarray*}
\sum_{r,s,u} \kappa^{ar} \kappa^{su}\left\{ \kappa_{rs}^{(u)} - \frac{1}{2}\kappa_{rsu}\right\} &=& \sum_{i=1}^n m_{1i} \sum_{r} \kappa^{ar} (r)_i\sum_{s,u} \kappa^{su} (s,u)_i \\
&&+ \sum_{i=1}^n n_{1i} \sum_{r,s,u} \kappa^{ar} \kappa^{su} \{(rs,u)_i - (ru,s)_i-(su,r)_i\}\\
&=&\sum_{i=1}^n m_{1i} \sum_r \kappa^{ar} (r)_i \delta_i^T(\tilde{X} K^{\beta\beta} \tilde{X}^T)\delta_i  \\
&&- \sum_{i=1}^n n_{1i} \sum_r \kappa^{ar} (r)_i \sum_{s,u} \kappa^{su} (su)_i\\
&=&\delta_a^T\sum_{i=1} K^{a\beta}\tilde{X}^T\delta_i m_{1i} \delta_i^T (\tilde{X} K^{\beta\beta} \tilde{X}^T)\delta_i\\
&&- \delta_a^T\sum_{i=1}^n n_{1i} K^{a\beta}\tilde{X}_i \delta_i F_i\\
&=& \delta_a^T K^{a\beta} \tilde{X}^T M_1 P_{\beta\beta} - \delta_a^T K^{a\beta} \tilde{X}^T N_1 F,
\end{eqnarray*}
where $K^{a\beta}$ is the matrix $K^{\beta\beta}$ if $a = 1,\ldots,k$ and is $K^{\beta\theta}$ if $a=k+1,\ldots,k+h$, further, if $a=k+1,\ldots,k+h$, we use the
abuse of notation $\delta_a$ to mean $\delta_{a-h}$, also, the matrices $P_{\beta\beta}$, $F_i$ and $F$ were defined in page \pageref{formulas}. Similarly, we obtain,
$$\sum_{R,s,u} \kappa^{aR} \kappa^{su}\left\{ \kappa_{Rs}^{(u)} - \frac{1}{2}\kappa_{Rsu}\right\} = \delta_a^T K^{a\theta} \tilde{Z}^T M_2 P_{\beta\beta} + \delta_a^T K^{a\theta}\tilde{Z}^T N_2F,$$
where, $K^{a\theta}$ is $K^{\beta\theta}$ if $a=1,\ldots,k$ and is $K^{\theta\theta}$ if $a=k+1,\ldots,k+h$. Further,
\begin{eqnarray*}
\sum_{r,S,u} \kappa^{ar} \kappa^{Su}\left\{ \kappa_{rS}^{(u)} - \frac{1}{2}\kappa_{rSu}\right\} &=& \sum_{i=1}^n m_{2i} \sum_{r} \kappa^{ar} (r)_i\sum_{S,u} \kappa^{Su} (S,u)_i \\
&&+ \sum_{i=1}^n n_{2i} \sum_{S,u} \kappa^{Su} (S)_i\sum_r \kappa^{ar}(ur)_i\\
&=&\sum_{i=1}^n m_{2i} \sum_r \kappa^{ar} (r)_i \delta_i^T(\tilde{X} K^{\beta\theta} \tilde{Z}^T)\delta_i  \\
&&+ \sum_{i=1}^n n_{2i} \sum_{S,u} \kappa^{Su} (S)_i \delta_a^T K^{a\beta}\tilde{X}_i \delta_u\\
&=&\delta_a^T\sum_{i=1} K^{a\beta}\tilde{X}^T\delta_i m_{2i} \delta_i^T (\tilde{X} K^{\beta\beta} \tilde{Z}^T)\delta_i\\
&&+ \delta_a^T\sum_{i=1}^n n_{2i}\sum_u K^{a\beta}\tilde{X}_i \delta_u \delta_u^TK^{\beta\theta}\tilde{Z}^T \delta_i\\
&=& \delta_a^T K^{a\beta} \tilde{X}^T M_2 P_{\beta\theta} + \delta_a^T K^{a\beta} L_1.
\end{eqnarray*}

In a similar fashion, we have
$$\sum_{r,s,U} \kappa^{ar} \kappa^{sU}\left\{ \kappa_{rs}^{(U)} - \frac{1}{2}\kappa_{rsU}\right\} = \delta_a^TK^{a\beta}\tilde{X}^TM_3 P_{\beta\theta} - \delta_a^T K^{a\beta}L_1,$$
where $P_{\beta\theta}$ was defined in page \pageref{formulas}. Now, analogously we have,
$$\sum_{R,S,u} \kappa^{aR} \kappa^{Su}\left\{ \kappa_{RS}^{(u)} - \frac{1}{2}\kappa_{RSu}\right\} = \delta_a^T K^{a\theta} \tilde{Z}^T M_4 P_{\beta\theta} - \delta_a^T K^{a\theta} L_2,$$
$$\sum_{R,s,U} \kappa^{aR} \kappa^{sU}\left\{ \kappa_{Rs}^{(U)} - \frac{1}{2}\kappa_{RsU}\right\} = \delta_a^T K^{a\theta} \tilde{Z}^T M_5 P_{\beta\theta} + \delta_a^T K^{a\theta} L_2,$$
$$\sum_{r,S,U} \kappa^{ar} \kappa^{SU}\left\{ \kappa_{rS}^{(U)} - \frac{1}{2}\kappa_{rSU}\right\} = \delta_a^T K^{a\beta} \tilde{X}^T M_5 P_{\theta\theta} + \delta_a^T K^{a\beta} \tilde{X}N_2G,$$
where $G$ was defined in page \pageref{formulas}. Finally, we have
$$\sum_{R,S,U} \kappa^{aR} \kappa^{SU}\left\{ \kappa_{RS}^{(U)} - \frac{1}{2}\kappa_{RSU}\right\} = \delta_a^T K^{a\theta} \tilde{Z}^T M_6 P_{\theta\theta} - \delta_a^T K^{a\theta} \tilde{Z}^T N_3 G.$$

\end{document}